\documentclass[referee]{aa}
\usepackage{graphics}

\begin{document}
\thesaurus{}%12.03.4;12.03.3;12.04.01;12.12.1;11.03.1} 

\title{A New Local Temperature Distribution Function for X--ray Clusters:  
Cosmological Applications}

\author{A.~Blanchard$^{1,2}$ \and R.~Sadat$^1$ \and J.G.~Bartlett$^1$ \and 
M.~Le~Dour$^1$}

\institute{ $^1$ Observatoire Midi-Pyr\'en\'ees, 14, Av. Edouard Belin, 31400 
        Toulouse, France\\
$^2$ Observatoire astronomique de Strasbourg, ULP,
            11, rue de l'Universit\'e, 67000 Strasbourg, France }

\offprints{A.Blanchard}
\mail{blanchard@astro.u-strasbg.fr}
   
\date{Received \rule{2.0cm}{0.01cm} ; accepted \rule{2.0cm}{0.01cm} }

\titlerunning{Cluster Temperature Function}
\authorrunning{Blanchard, Sadat, Bartlett \& Le Dour }
\maketitle
 
\begin{abstract} 

We present a new determination of the local temperature 
function of X-ray clusters using a sample of X-ray clusters with fluxes above 2.2
$10^{-11}$erg/s/cm$^2$ in the $[0.1-2.4]$ keV band, most of these clusters come from the Abell XBAC's 
sample to which a handfull of known non-Abell clusters
 has been added. We estimate this sample to be 85\% complete, and should 
therefore provide an useful estimation of  the present-day  
number density of clusters.
Comprising fifty clusters for which the temperature information is available, it is the largest complete sample of this kind. It is 
therefore expected to significantly improve the 
estimation of the temperature distribution function of clusters. We find that the resulting temperature function is higher than 
previous estimations, but it agrees with the temperature distribution 
function inferred from the BCS and RASS luminosity function 
(Ebeling et al., 1997; De Grandi et al. 1999a).  We have used this sample to constrain the amplitude of the matter 
fluctuations $\sigma_c $ on cluster's scale of $8\sqrt[3]{\Omega_0}^{-1}h^{-1}$Mpc, 
assuming a mass-temperature relation based on recent numerical simulations. 
We find $\sigma_c \sim 0.6\pm 0.02$ for an $\Omega_0 = 1$ model (for which  
$\sigma_c = \sigma_8$). Our sample 
provides an useful reference at $z \sim 0$ to use in the application of the 
cosmological test based on the evolution of X-ray clusters abundance 
(Oukbir \& Blanchard 1992, 1997). We have therefore 
estimated the temperature distribution function at $z = 0.33$ using  
Henry's sample of 
high-z X-ray clusters (Henry, 1997; hereafter H97) and performed a preliminary
estimate of $\Omega_0$. We find that the abundance of clusters at $z = 0.33$ 
is significantly smaller, by a factor larger than 2, which shows that the 
EMSS sample provides strong evidence for evolution of the cluster 
abundance.  A likelihood analysis leads to a rather high value of the mean 
density parameter of the universe: $\Omega_0 = 0.92^{+0.255}_{-0.215}$ ($1 \sigma$) for  open universes and 
$\Omega_0 = 0.865^{+0.35}_{-0.245}$  for flat 
universes, which is consistent with a
previous independent estimation based on the full EMSS sample by 
Sadat et al.(1998). Some systematic uncertainties which could alter 
this result are briefly discussed. 

\end{abstract}  

\section {Introduction} 
 
Clusters are believed to be the largest virialized 
concentrations of dark matter. Therefore, they offer privileged regions for 
studying dark matter distribution on large scales in the universe.
X-ray and weak lensing mass measurements have been added to 
the traditional mass estimates based on optical velocity dispersions. 
Weak lensing analyses are still in their infancy and at present
there exists no sample of weak lensing observations large enough to establish 
a mass function. The velocity distribution function of clusters 
could be used to derive the mass function. However, as 
Evrard pointed out (Evrard, 1989) the error on individual measurements 
can introduce a significant overestimate. Furthermore, velocity measurements 
at least in the case of distant clusters, can be corrupted by 
projection effects that might be difficult to remove in practice 
(Frenk et al., 1990). Moreover, Sadat et al. (1998) have shown that 
X-ray temperatures of some of the CNOC clusters show a significant 
difference with what is expected from their velocity dispersion measurements. 
For these reasons, it has been argued that the X-ray temperature 
is a better indicator of cluster mass.  Numerical simulations have 
greatly helped to understand the relation between X--ray temperature and 
the mass, and useful constraints have been placed on the amplitude and 
the shape of the spectrum of mass density
fluctuations. Still, the size of the samples of X-ray clusters 
homogeneously selected is limited: 25 clusters in the 
Henry and Arnaud (1991, hereafter HA91) sample and 30 clusters in Markevitch's
(1998). ROSAT selected clusters samples have significantly improved the 
situation in this domain. In order to achieve stronger constraints on 
theoretical models, it will be necessary to obtain more temperature 
measurements, a vast program that will probably be possible 
with the next generation of X-ray satellites such as AXAF and XMM. 

The cluster X--ray temperature function is a powerful 
tool for cosmology. Provided that the mass-temperature relation is reasonably 
well known, the Press and Schechter (1974) formalism allows one to constrain 
the amplitude and the shape of the power spectrum for a given cosmological 
background density (see Bartlett, 1997, for a recent review on the subject).
Since there is a nearly complete degeneracy between the amplitude of the 
fluctuations and the mean density of the universe, the cosmological parameter 
$\Omega_0$ cannot be determined solely from the local temperature function.
However, the evolution of this temperature distribution function, once 
normalized to the present day cluster abundance, varies significantly 
with $\Omega_0$ offering an interesting new cosmological probe (Oukbir \& Blanchard 1992, hereafter OB92; 
Hattori \&  Matsuzawa, 1995). This test has received considerable 
attention in recent years (Donahue, 1996; Carlberg et al., 1997; 
H97; Oukbir \& Blanchard 1997, OB97, hereafter; Blanchard \& Bartlett, 1998; Eke et al., 1996, 1998; 
Sadat et al., 1998; Viana \& Liddle, 1996, 1999a; Blanchard et al., 1999; 
Donahue et al., 1999; Donahue and Voit, 1999; Reichart et al., 1999).  Variants have been proposed using the 
Sunyaev--Zeldovich (Barbosa et al, 1996) and weak lensing effects 
(Kruse \& Schneider, 1999). Modeling the redshift distribution of the 
EMSS X-ray selected sample (OB97) given the absence of 
any significant evolution in the 
$L_x-T_x$ relation (Mushotsky \& Sharf, 1997; Sadat et al., 1998), 
seems to favor high value of the density parameter (Sadat et al, 1998; 
Reichart et al, 1999). Modeling the RDCS redshift distribution 
leads to consistent  results (Borgani et al, 1998). 
A more direct estimate, free from any consideration on the possible 
evolution in the $L_x-T_x$ relation, could be obtained from the measurement 
of the evolution of the temperature distribution function, which in turn requires a good knowledge of the selection function of the sample of clusters. 
The first sample of X-ray selected clusters at non--zero redshift
with measured temperatures has recently become available (H97) and has 
led to an apparent median value of $\Omega_0$
in the range $0.2-0.6$ (H97; Eke et al, 1998), although higher values 
were found by Viana and Liddle (1999a) and Blanchard et al. (1999). It has been 
argued that current data are not good enough to allow a reliable estimate 
of the mean density of the universe from such techniques 
(Colafrancesco et al., 1997), a conclusion that 
appears wise given that the local sample of X-ray clusters used up to now 
is that of HA91 which contains only 25 clusters while the high--redshift 
sample comprises only 10 clusters with moderately accurate temperature 
measurements (recently, the redshift of one cluster in the sample has been 
revised, reducing the sample to 9 clusters, Donahue et al., 1999). However,
the fact that some conclusions have already been drawn, even if too optimistic 
regarding possible systematics, demonstrates the power of this test: 
clearly ten more clusters or so at high 
redshift and an accurate determination of the local temperature distribution 
function would provide a very robust determination of the 
mean density of the universe. In fact, going to high redshift makes a dramatic 
difference in the abundance of hot clusters (OB92), and 
the existence of few EMSS clusters with a high temperature has been argued to 
already provide a strong evidence for a low density universe (Donahue 1996; 
Bahcall \& Fan, 1998; Donahue et al., 1998;  Eke et al., 1998).\\

Until recently, the X-ray cluster temperature distribution function was
inferred from catalogs built from the HEAO1 survey in the 2-10 keV band. 
The need for a new estimation of the temperature distribution function 
has been recognized. Markevitch (1998, hereafter M98) has provided such a new estimate based on a sample 
of X-ray clusters for which ROSAT fluxes were available. 
Most of his clusters come from the XBACS sample of bright Abell 
clusters (Ebeling et al., 1996). Temperatures are 
derived from ASCA observations, and for this reason clusters at redshift 
smaller than 0.04 were not considered.  This means that the sample is 
not a strictly flux limited sample.  M98 corrected both the fluxes 
and temperatures of his sample for the presence of cooling flows in the 
central regions. Cooling flows are believed to be an important source of 
dispersion in the $L_x-T_x$ relation (Fabian et al., 1994; M98; 
Arnaud \& Evrard, 1999), but have negligible effect on the temperature 
distribution function. \\

In the following, we  present the analysis of a new sample containing 
fifty clusters essentially based on the XBACS sample (Ebeling et al., 1996), 
using a flux limit similar to that of M98.  Note that M98 considered his 
sample to be reasonably complete. In the present analysis, the effect of 
temperature error measurements is explicitly taken into account using a 
Bayesian correction. Our sample does not require any correction for 
redshift incompleteness. We also do not correct for cooling flows in order 
to allow a direct comparison between our sample and those at high redshift. 
The resulting temperature distribution function we obtain is smooth 
and can be used for useful comparison with theoretical models. 
We have performed such a comparison in order to constrain 
the density fluctuation power spectrum with different values of the 
cosmological background density $\Omega_0$.
This paper is organized as follows: section 2 summarizes the results from 
previous estimates of the local temperature function. In section 3 
we give a brief description of our sample. In section 4, we present our   
method to calculate the local temperature function as well as 
a determination of the distribution function of the cluster abundance 
estimator. We also discuss how our temperature function compares with previous 
results. Section 5 presents the first application of this new temperature 
distribution function to constrain cosmological parameters.  

\section{Previous estimates} 

\begin{figure*}
\resizebox{\hsize}{!}{\includegraphics{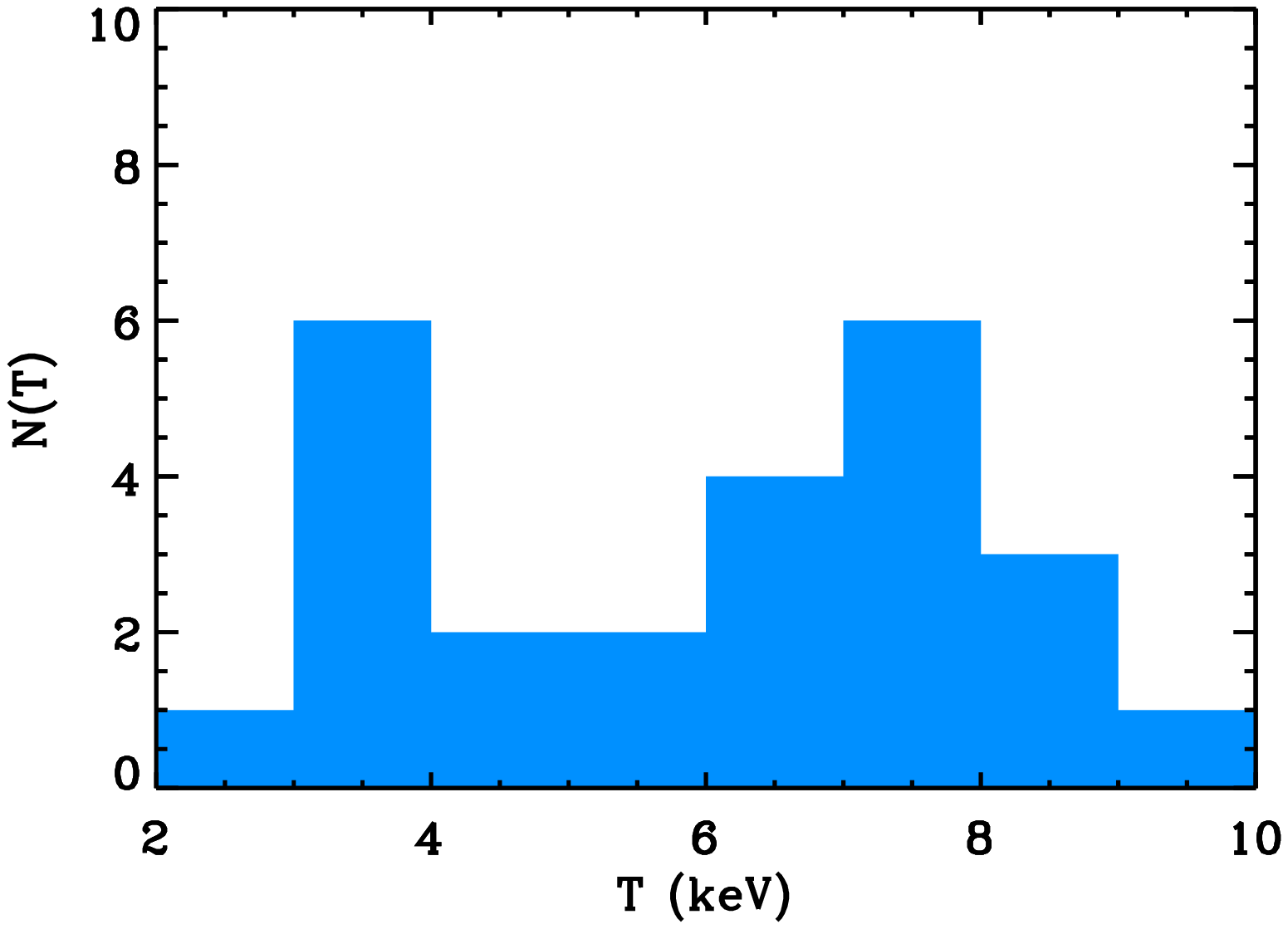}\includegraphics{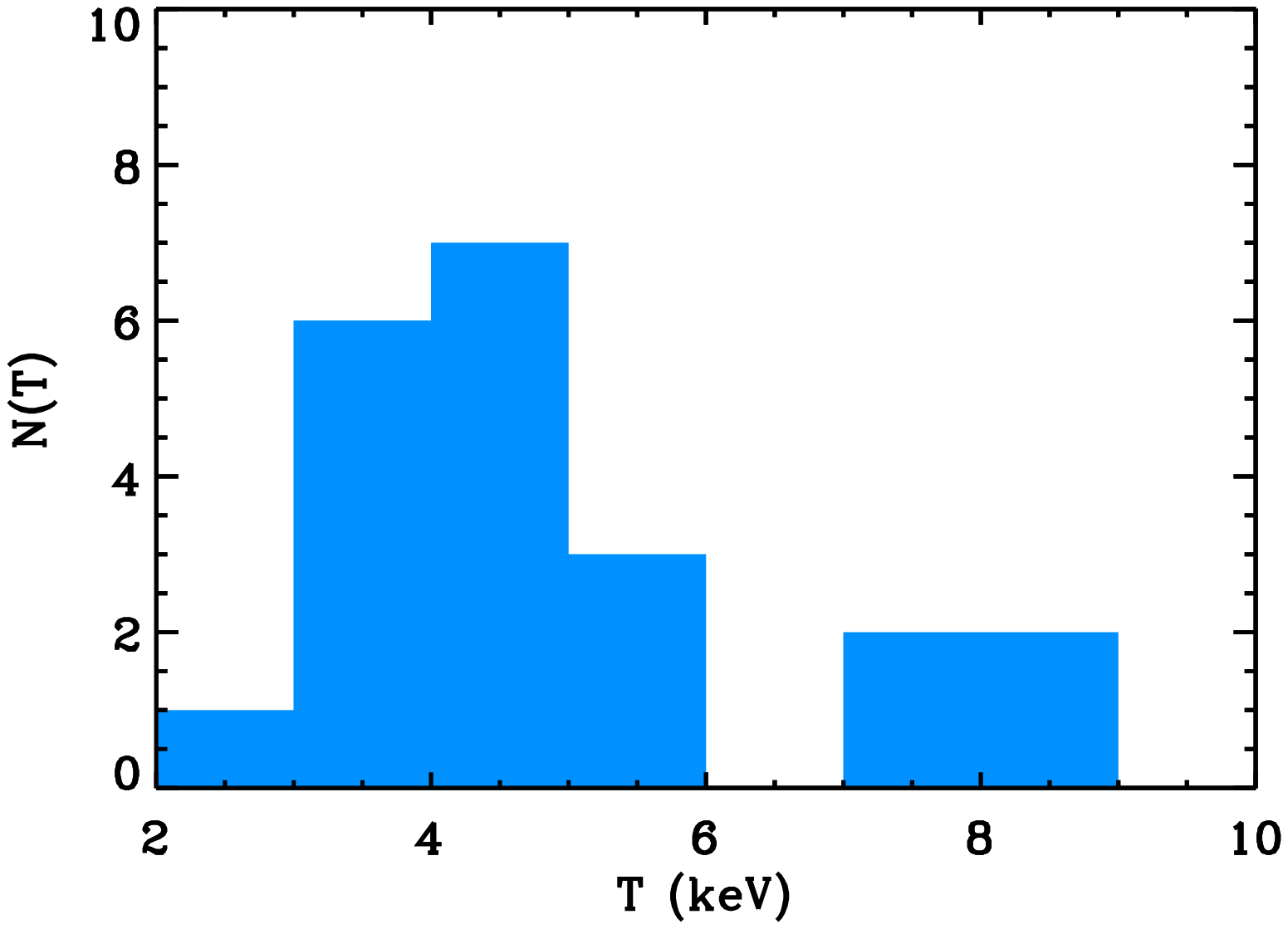}}
\hspace*{1.8cm}a) \hspace*{8.4cm} b)
\caption{\label{fig:histoT1} The distribution of clusters with revised 
temperatures in the Henry \& Arnaud sample, HA91 (left panel). The right 
panel shows the histogram of the 21 clusters in the Edge et al. (1990) sample,
not included in HA91. 
%Because both samples are considered as X-ray selected, one should 
%expect the same distribution. Clearly this is not the case. The reason for 
%this difference is unclear, but translates in a significant difference 
%in the inferred temperature function.
}
\end{figure*}

In order to determine the temperature function, it is important to have a 
well controlled sample of clusters. The knowledge of the selection 
function is therefore critical. For instance, the completeness of the Edge et al. 
(1990) sample (hereafter E90) was 
questioned by HA91, and in fact the difference between the 
E90 and HA91 temperature functions is striking. We therefore re--examine 
these two samples, although the revised version of HA91 appears to be closer 
to E90.  Rather than plotting the temperature function, we directly 
examine the histogram of the number of clusters in temperature bins 
for each sample.  In Figure \ref{fig:histoT1}a, we show the histogram
of the revised temperatures in the HA91 sample, while in Figure 
\ref{fig:histoT1}b, we give the corresponding histogram for the E90 
clusters {\em not present} in the HA91 sample; this latter will be referred to 
as the E-HA sample. Comparison of the two histograms clearly reveals 
a striking difference: there is an apparent deficit of clusters 
with $T \sim 4-6$ keV in the HA sample as compared to E-HA. 
Obviously, the smaller number of 4 keV clusters present in the HA sample 
leads to a lower number density. This is of critical importance in the 
theoretical modeling, because the abundance of 4 keV clusters is 
crucial in determining the amplitude of the fluctuations on a 8$h^{-1}$Mpc 
scale (for $\Omega_0 = 1$). The E-HA set is astonishingly different: 
the number of clusters reaches a maximum at $T = 4$ keV, and then tends to 
decrease monotonically for higher $T$, meaning that there are relatively 
fewer hot clusters in the E-HA data set (4 clusters from the E90 sample
with poor temperature measurements are not taken into account in E-HA, 
which adds 3 clusters to the region $T > 8$ keV, but this does not change 
the shape of the histogram). Although 
statistical analysis of such a small sample is a
hazardous exercise, this difference is highly significant: assuming that the number of clusters found in the HA sample between 4 and 6 keV (4 clusters), the probability to have ten clusters or more in the E-HA sample in the same range is less than $10^{-3}$. When dealing with small samples, one must of course be careful of
spurious noise introduced by Poisson statistics. However, using different tests
we always find this difference  significant at a confidence level greater than 95\%. It is difficult to understand the 
origin of such a difference: both samples are X-ray selected
and incompleteness is expected to be nearly independent of temperature.

It is possible that the existence of 
large--scale fluctuations, as can be seen from 
the visual appearance of the clusters 
distribution or from the long--range correlation function, indicates that 
noise due to a finite number of objects may be larger than that
expected from Poisson noise. 

The comparison we have presented in Fig. \ref{fig:histoT1} clearly 
calls for a larger, complete flux limited sample strictly X-ray selected. 
However, only the Rosat All Sky Survey (RASS) would allow the construction of such a sample. The Bright Cluster Survey BCS (Ebeling et al., 1998) and the RASS South (De Grandi et al., 1999b) have just become available at the time 
of writing, but the lack of temperature information for a significant 
fraction of these clusters prevents us from using these samples
to usefully estimate the temperature distribution function. 
Recently, M98 provided a new local temperature function based on the 
XBACS sample (Ebeling et al, 1996). 
XBACS is an essentially complete, flux-limited X-ray sample of Abell clusters.
In principle, the restriction to Abell clusters implies that the sample 
is possibly incomplete.  There are three possible sources of incompleteness:
i) low mass (i.e., low temperature) clusters may be
missing because they are not rich enough to be selected 
by the Abell criteria; ii) distant clusters could be missed because at 
large redshifts clusters hardly meet the Abell criteria; 
iii) there is a significant difference between the optical
based Abell criteria and those based on X-ray fluxes.
M98 restricted his sample to the redshift range 
$0.04\leq  z \leq 0.09$, the lower limit being imposed by ASCA. 
This redshift restriction eliminates intrinsically faint clusters from 
his sample. M98 corrected for this incompleteness 
by using a weight to compensate for the selection function. 
The resulting temperature function is noticeably higher than previous 
estimates, although it is claimed to be consistent with them. 
However, the correction procedure could introduce a systematic error 
in the derived abundance of cool clusters as we will see in the next 
section.

\section{The present sample} 

\begin{center}

\begin{table*}[ht]
   \begin{tabular}{llllll}
      \hline
     \hline
Cluster & z &$T_{X}$$^{1}$ & $L_{X}$$^{2}$ & $s_{X}$$^{3}$ &  Ref$^{4}$.\\
      \hline
      Virgo&   0.0036 &      2.2$^{    +0.1 }_{    -0.1 }$  &
     1 &      82.0 &{\sl 6,10 } \\
      A1060&    0.0114&      3.25$^{     +0.1 }_{    -0.1 }$  &
     0.47&      7.&{\sl 4,1 } \\
     A3526&    0.0109&      3.54$^{     +0.1}_{    -0.1}$  &
     0.83&      19.&{\sl 8,1} \\
     A262 &    0.0161 &      2.15$^{    + 0.1}_{    -0.1}$  &
     0.55&      4.8&{\sl 4,1 } \\
     A3581&    0.02&      2.0$^{    + 0.1}_{    -0.1}$  &
     0.57&      2.9&{\sl 5,1 } \\
     A1367&    0.0215 &      3.55$^{     +0.1}_{    -0.1}$  &
      1.63 &      8.3&{\sl 4,1 } \\
     A1656&    0.0232 &      8.2$^{     +0.1}_{    -0.1}$  &
      7.21 &      31.6 &{\sl 9,1 } \\
     A4038&    0.029&      3.3$^{    + 2.6}_{    -1.3}$  &
      1.9&      5.27 &{\sl 2,1 } \\
     A2199&    0.0309 &      4.5$^{     +0.1}_{    -0.1}$  &
      3.65 &      9.5&{\sl 7,1 } \\
     A496&    0.032&      4.1$^{     +0.1}_{    -0.1}$  &
      3.54 &      7.5&{\sl 4,1 } \\
     A2634&    0.0321 &      3.7$^{    + 0.3}_{    -0.3}$  &
     0.94&      2.3&{\sl 4,1 } \\
     A2063&    0.0337 &      3.7$^{    + 0.1}_{    -0.1}$  &
      2.03 &      3.7&{\sl 4,1} \\
     A2052&    0.0348 &      3.1$^{    + 0.3}_{    -0.3}$  &
      2.52 &      4.7&{\sl 2,1 } \\
    Cl0336&    0.0349 &      3.$^{    +0.1}_{    -0.1}$  &
      5.87 &      8.76 &{\sl 4,1 } \\
     A2147&    0.0356 &      4.9$^{     +0.3}_{    -0.3}$  &
      2.84 &      5.3&{\sl 4,1 } \\
    A576&    0.0381 &      4.3$^{    + 0.5}_{    -0.4}$  &
      1.39 &      2.24 &{\sl 2,1 } \\
     0422--09&    0.039 &      2.9$^{    + 0.5}_{    -0.4}$  &
      {\em 2.$^*$} &      {\em 3. $^*$} & {\sl 2,2 } \\
     A3571&    0.0391 &      6.9$^{    + 0.3}_{    -0.3}$  &
      7.36 &      10.9 &{\sl 3,1 } \\
    A2657&    0.04&      3.7$^{    + 0.3}_{    -0.3}$  &
      1.6&      2.33 &{\sl 3,1 } \\
     A2589&    0.0416 &      3.7$^{    + 0.6}_{    -0.6}$  &
      1.86 &      2.5&{\sl 6,1 } \\
     A119&    0.044&      5.8$^{    + 0.4}_{    -0.4}$  &
      3.23 &      3.8&{\sl 3,1 } \\
    MKW3s&    0.045&      3.5$^{    + 0.2}_{    -0.2}$  &
      3.&      3.43 &{\sl 3,3 } \\
     A1736&    0.0461 &      3.5$^{    + 0.4}_{    -0.4}$  &
      2.37 &      2.6&{\sl 3,1 } \\
    A3376&    0.0464 &      4.3$^{    + 0.6}_{    -0.6}$  &
      2.48 &      2.69 &{\sl 3,1 } \\
     A3558&    0.0478 &      5.5$^{    + 0.3}_{    -0.3}$  &
      6.27 &      6.2&{\sl 3,1 } \\
\hline
 \hline
\end{tabular}
\hspace*{2mm}
   \begin{tabular}{llllll}
      \hline
     \hline
Cluster & z &$T_{X}$$^{1}$ & $L_{X}$$^{2}$ & $s_{X}$$^{3}$ &  Ref$^{4}$.\\
      \hline
   A1644&    0.048&      4.7$^{     +0.8}_{    -0.8}$  &
      3.52 &      3.64 &{\sl 2,1 } \\
    A4059&    0.048&      4.0$^{    + 0.1}_{    -0.1}$  &
      3.09 &      3.12 &{\sl 4,1 } \\
    A3562&    0.0499 &      3.8$^{    + 0.8}_{    -0.8}$  &
      3.33 &      3.08 &{\sl 2,1 } \\
    A3395&    0.05&      4.8$^{     +0.4}_{    -0.4}$  &
      2.8&      2.65 &{\sl 3,1 } \\
     A85&    0.0518 &      6.1$^{     +0.2}_{    -0.2}$  &
      8.38 &      7.2&{\sl 3,1 } \\
     A3667&    0.053&      7.$^{    + 0.6}_{    -0.6}$  &
      8.76 &      7.3&{\sl 3,1 } \\
     A754&    0.0534 &      7.6$^{     +0.3}_{    -0.3}$  &
      8.01 &      12.&{\sl 6,1 } \\
     A780&    0.05384 &      3.8$^{     +0.2}_{    -0.2}$  &
      6.63 &      4.8&{\sl 3,1 } \\
    A3391&    0.054&      5.7$^{    + 0.7}_{    -0.7}$  &
      2.32 &      2.39 &{\sl 3,3  } \\
    A3158&    0.059&      5.5$^{     +0.6}_{    -0.6}$  &
      5.31 &      3.57 &{\sl 2,1 } \\
     A3266&    0.0594 &      7.7$^{     + .8}_{     -0.8}$  &
      6.15 &      4.8&{\sl 3,1 } \\
     A2256&    0.0601 &      7.5$^{     +0.4}_{    -0.4}$  &
      7.05 &      4.9&{\sl 3,1 } \\
    A133&    0.0604 &      3.9$^{    + 1.6}_{    -0.7}$  &
      3.57 &      2.29 &{\sl 2,1 } \\
      A1795&    0.0616 &      6.$^{    + 0.3}_{    -0.3}$  &
      11.1 &      6.7&{\sl 3,1 } \\
     A3112&    0.07&      4.7$^{    + 0.4}_{    -0.4}$  &
      7.7&      3.6&{\sl 3,1 } \\
      A399&    0.0715 &      7.4$^{    + 0.7}_{    -0.7}$  &
      6.45 &      2.9&{\sl 3,1 } \\
     A2065&    0.072&      5.4$^{    + 0.3}_{    -0.3}$  &
      4.95 &      2.2&{\sl 3,1 } \\
    A401&    0.0748 &      8.3$^{    + 0.5}_{    -0.5}$  &
      9.88 &      4.26 &{\sl 3,1 } \\
    A2029&    0.0767 &      8.7$^{    + 0.3}_{    -0.3}$  &
      15.35 &      6.16 &{\sl 3,1 } \\
     A1651&    0.0825 &      6.3$^{    + 0.5}_{    -0.5}$  &
      8.25 &      2.7&{\sl 3,1 } \\
    A1650&    0.085&      5.6$^{    + 0.6}_{    -0.6}$  &
      7.81 &      2.56 &{\sl 3,1 } \\
    A2597&    0.085&      3.6$^{    + 0.2}_{    -0.2}$  &
      7.97 &      2.59 &{\sl 3,1 } \\
    A2142&    0.0899 &      9.$^{    + 0.2}_{    -0.2}$  &
      20.74 &      6.14 &{\sl 11,1 } \\
     A478&    0.09&      7.1$^{     +0.4}_{    -0.4}$  &
      12.95 &      3.9&{\sl 3,1 } \\
    A2244&    0.097&      7.1$^{      +1.2}_{     -1.2}$  &
      9.09 &      2.28 &{\sl 2,1 } \\
\hline
       \hline
          \end{tabular}

\noindent $^{1}$ in {\sl keV}\\
$^{2}$ in 10$^{+44}$erg/s ($h = 0.5$)\\
$^{3}$ in 10$^{-11}$erg/s/cm$^2$\\
$^{4}$ the first reference is for temperature, the second for the luminosity\\
$^*$ inferred from a non-ROSAT measurement.\\
                \caption{ \label{Table 1}  X-ray temperatures, 90\% uncertainties,  ROSAT 0.1--2.4 keV luminosities (h$ = 0.5$) and fluxes.}
${\bf Ref.}$
{\sl (1)} Ebeling et al. 1996;
{\sl (2)} David et al., 1993;
{\sl (3)} Markevitch, 1998;
{\sl (4)} Fukazawa et al., 1998 (temperature are given with the emission from the central $0.05h^{-1}$Mpc excluded);
{\sl (5)} Johnstone et al., 1998;
{\sl (6)} Arnaud \& Evrard, 1999;
{\sl (7)} Arnaud, 1994;
{\sl (8)} Yamashita et al., 1992;
{\sl (9)} Hughes et al., 1993;
{\sl (10)} Ebeling et al., 1998;
{\sl (11)} White et al., 1994;
\end{table*} 
\end{center}

As it is clearly an advantage to have a survey which is not redshift 
restricted, we have collected existing information on clusters with 
$ f_x \geq 2.2 10^{-11}$ erg/s/cm$^{2}$ in the ROSAT 
band 0.1--2.4 keV (and with $| b_{II}| >  20^\circ $).  In practice, most 
of the clusters were obtained from  the XBACS sample complemented by a handful of non-Abell clusters which satisfy our flux requirement
(MKW3s, Virgo, Cl0336 and 0422--09). For all these clusters, 
temperature measurements were found  in the literature.
In the following analysis, the redshift range is still restricted to $z< 0.1$, because of 
the possible substantial incompleteness of the sample at higher redshifts. 
Our final sample contains fifty clusters. It is important to examine its completeness.

\begin{figure}
\resizebox{\hsize}{!}{\includegraphics{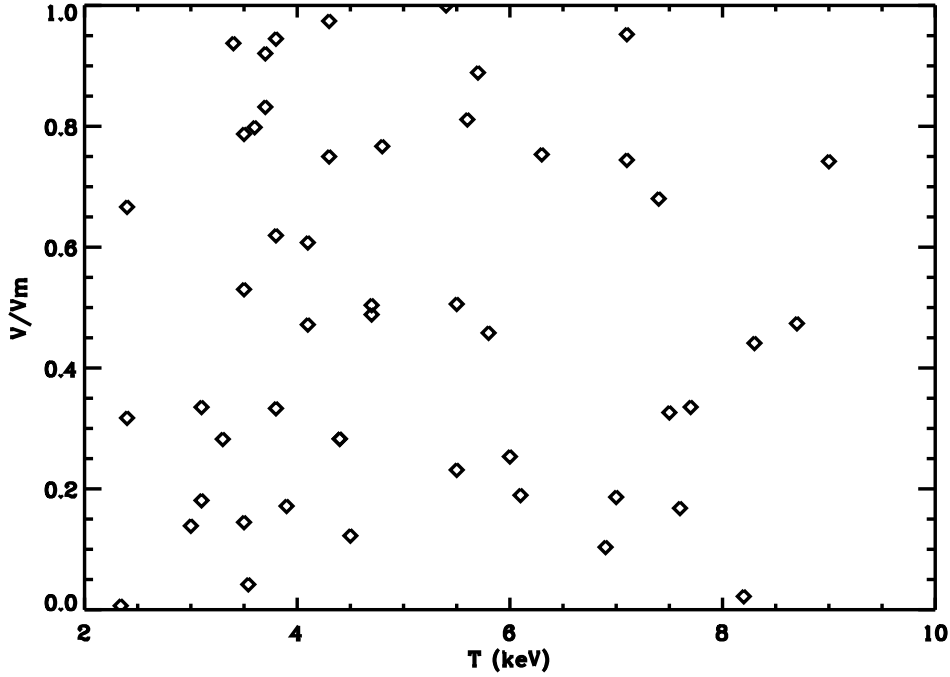}}
\caption{ \label{fig:vvm} $V/V_m$ distribution for clusters in our 
sample as a function of temperature. No sign of 
incompleteness can be seen from this plot for clusters with 
$T \geq 3.5$keV.}
\end{figure}

A classical test for checking the homogeneity of a survey is to use the 
$V/V_m$ test (Schmidt, 1968). Applying this test, we find a mean value 
of $V/V_m$ = 0.490, consistent with a homogenous sample. 
The plot of $V/V_m$ versus temperature given in Figure \ref{fig:vvm} 
does not reveal any unexpected trends. 
The recent availability of the BCS and RASS samples allows us to check this
in a more quantitative way: at our chosen flux limit, four clusters
in the BCS and RASS1 are not present in our sample (RXJ0419, IIZw108, Z5029, NGC1550). None of these 
(non-Abell) clusters has a measured temperature, but from their 
luminosity, only two of them would have a temperature 
greater than 4 keV. This is consistent with our sample being complete at a 
level of 95\% for $T \geq 4$ keV, while the 
completeness of the BCS and RASS is estimated to be better than 90\%.
Therefore, our estimated completeness is 85\% and we will treat it as 
an essentially complete, X-ray selected ensemble of clusters, at least for clusters with temperature greater 
than 4 keV. It is the largest homogenous sample of X-ray selected clusters 
currently available for the estimation of the temperature 
distribution function. Notice that the possible incompleteness of our sample 
would imply that the inferred temperature function could be an 
{\em under-estimate}. 

\begin{figure*}
\resizebox{\hsize}{!}{\includegraphics{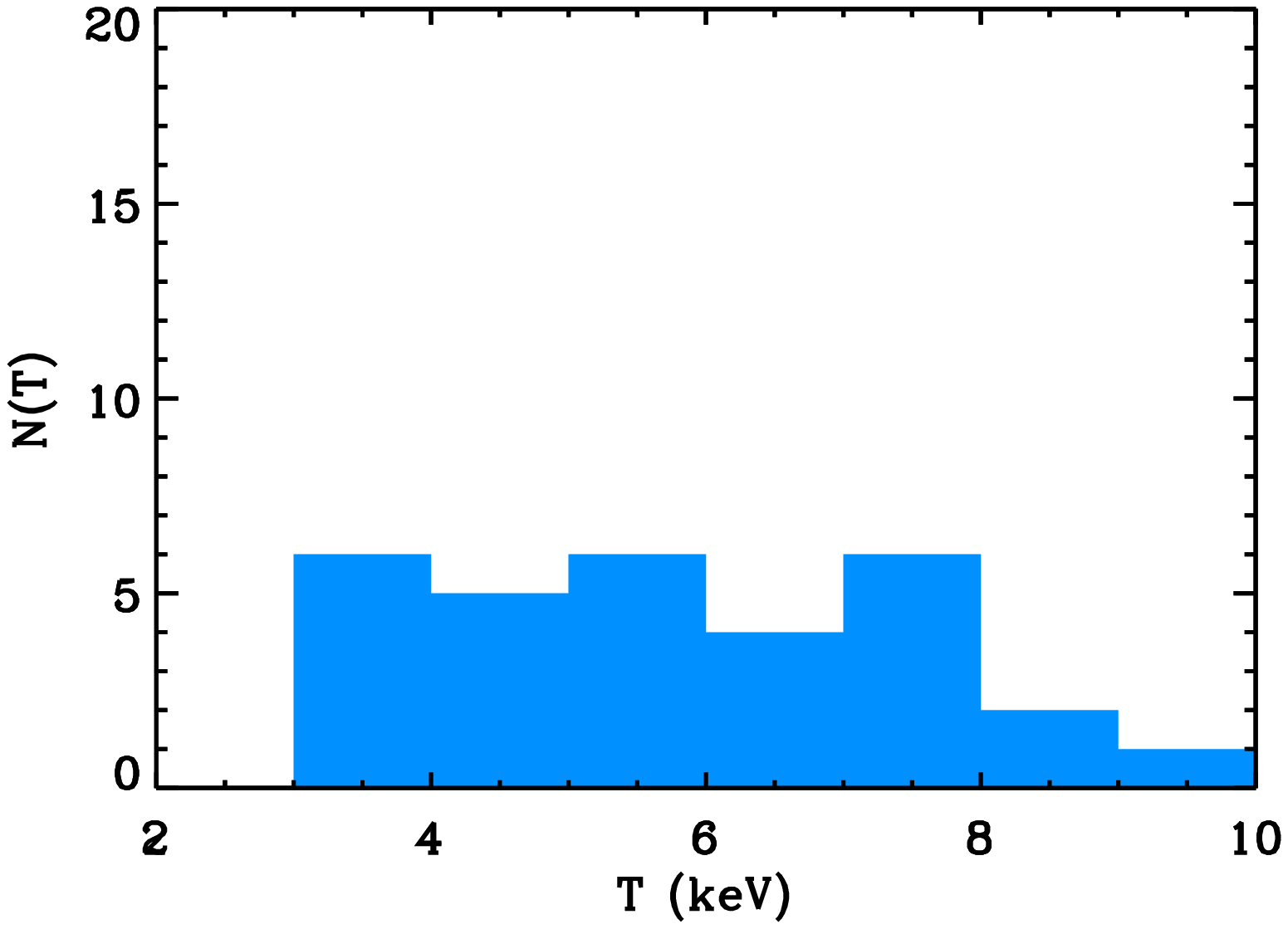}\includegraphics{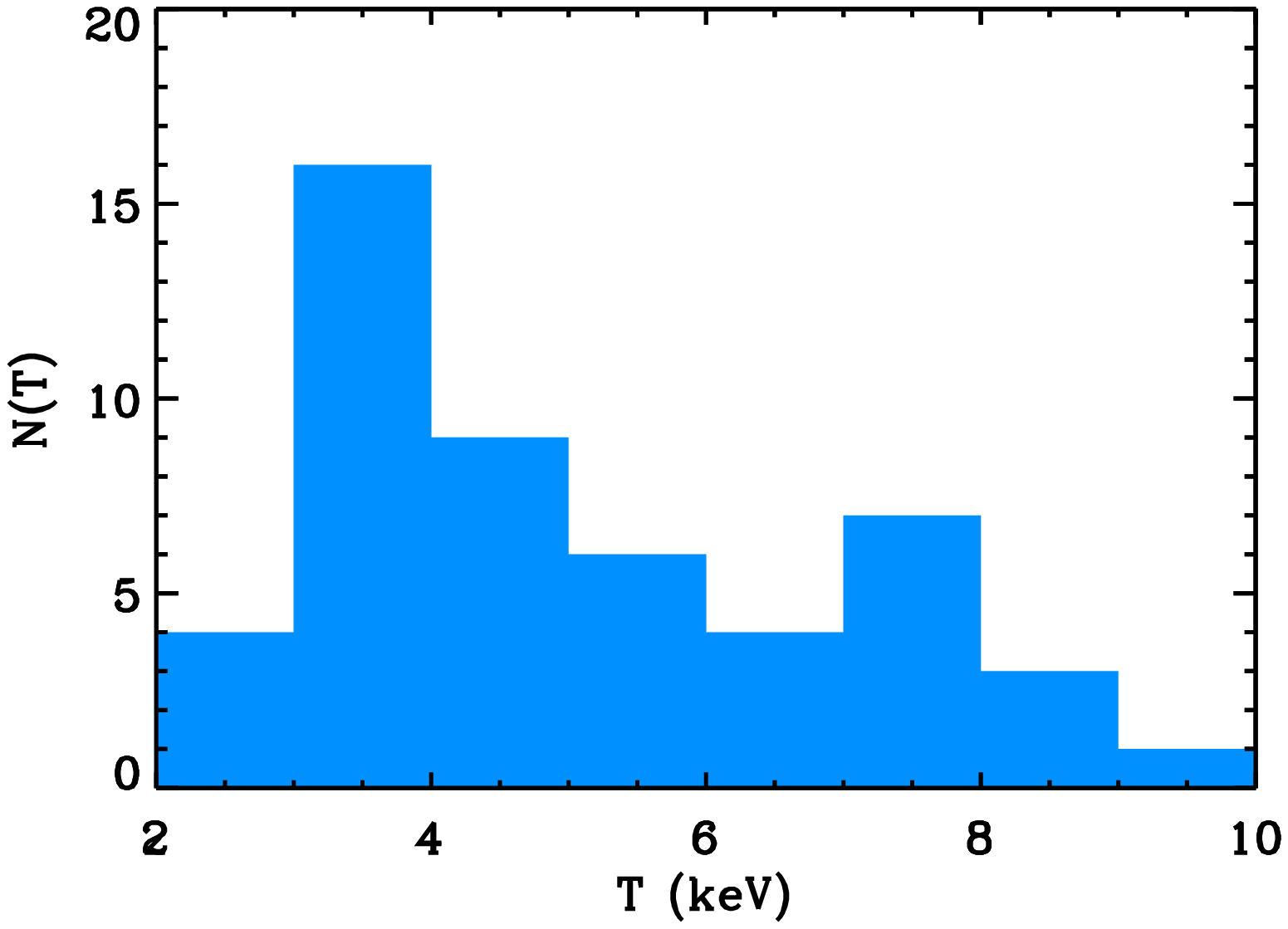}}
\hspace*{1.8cm}a) \hspace*{8.4cm} b)
\caption{ \label{fig:histoT2} Same as in Fig. \ref{fig:histoT1} for clusters 
selected in the ROSAT band. 
Figure a) represents the distribution of clusters in Markevitch's 
 sample. b) represents the 
histogram 
of the 50 clusters in our sample (XBACS clusters with flux 
greater than $2.2 \; 10^{-11}$erg/s/cm$^2$ and redshift $z < 0.1$). 
The shapes clearly differ for temperature smaller than 5 keV, 
while they are nearly identical above. This difference is mainly due to the fact that 
the Markevitch sample is restricted to redshift $> 0.04$. 
}
\end{figure*}

        At redshifts $z \geq 0.04$, our sample is similar to M98.
The temperature histograms of both M98 sample and ours are shown 
in Figure \ref{fig:histoT2}. As one can see, the two histograms are 
nearly identical on the hot end, but differ on the cool end
(for $T < 5$ keV). This difference is not surprising given the
restricted redshift range investigated by M98. However, it is important
to investigate the temperature function on the cool end since the number of 
clusters in the two samples differs noticeably. Although M98 corrected 
for the incompleteness of his sample, using a larger sample will result in a 
statistically better estimate with reduced sensitivity to 
systematics. 
Such a sample has two main advantages: with a selection in the ROSAT band, 
we are much less sensitive to the selection function of the 2.--10. keV band, 
which obviously favors hot clusters (even if in principle the effect of 
the selection function can be corrected for), and the number of clusters in 
our sample is twice as large as the HA91 sample, leading to better
statistics especially on the cool end ($T \sim 4$ keV). 
We have examined the temperature histograms for 
clusters with fluxes in the range  $ 2.2 -4.0 \; 10^{-11}$ erg/s/cm$^{2}$
and those with $f_{X} > 4.0 \;10^{-11}$ erg/s/cm$^{2}$. 
We find that these two histograms differ noticeably in the same sense 
as the HA91 and E-HA samples as discussed previously.
The reason of this difference is unclear, but it is clearly not due 
to the 2.--10 keV selection only. Inspection of the redshift distribution does 
not provide evidence that this could be due to large--scale structures
(although the BCS reveals significant fluctuations around $z \sim 0.05$).\\

\section{The Determination of the Temperature Function} 
\subsection{The method}
Estimation of the luminosity distribution function $\phi(L)$ from a sample 
for which the selection function is well known is relatively straightforward:
the observed number of clusters in the sample $\cal N$ in the luminosity range
$(L-L+\Delta L)$ (with $\Delta L \ll L$)
is the realization of a stochastic Poisson process with a mean value 
$ \phi(L) V(L)\Delta L$, where $V(L)$ is the volume searched  
for objects with luminosity $L$. When the selection function $s(z)$ is that 
of a strictly flux limited sample, $V(L)$ is just the volume 
$V_m$ corresponding to 
the maximum distance at which the cluster would have been detected. If the 
selection function $s(z)$ is more complex, then this volume can always be 
computed as an integral:
\begin{displaymath}
V(L) = \int_0^{+\infty} s(z) dV(z)
\end{displaymath}

An unbiased estimation of $ \phi(L) \Delta L$  is therefore given by 
${\cal N}/V(L) $. In the following, when $B$ is an estimator of some 
quantity $A$, we use the notation below:
\begin{displaymath}
A \equiv B
\end{displaymath}
According to this notation:
\begin{displaymath}
\phi(L) \Delta L \equiv {\cal N}/V(L) = \sum 1/V
\end{displaymath}
where the summation is performed over all the relevant clusters in the sample.
This immediately provides an estimator of the integrated luminosity function:
\begin{displaymath}
\phi(>L)  \equiv \sum_{L_i \geq L} 1/V(L_i)
\end{displaymath}
We can now recover the standard estimator of the temperature distribution 
function: 
\begin{equation}
N(T) \Delta T = \int_0^{+\infty}\phi(L,T) \Delta T dL  \equiv  \sum_{i} 1/V(T_i)
\label{eq:nesT}
\end{equation}
where the summation is performed over all clusters in the sample in the 
relevant temperature range. It is important to notice that this estimator is  
unbiased and fully accounts for the intrinsic dispersion in the 
temperature-luminosity relation. This procedure may not always be used in 
practice, if for some region of the parameter space in the sample $V(L) = 0$. 
This is the case for instance, in a flux limited sample which is 
further restricted to a redshift range. Because of this further restriction, 
clusters fainter than some intrinsic luminosity may simply not be present in 
the sample, consequently leading to $V(L) = 0$. In this case, the estimation 
from Eq. (\ref{eq:nesT}) is clearly not appropriate anymore. 
The contribution of the missing clusters can 
however be estimated, provided the probability distribution 
of clusters at a given temperature $p(L|T)$ is known. The observed number of 
clusters is still a
Poisson realization of a process whose mean is given by:
\begin{eqnarray}
\nonumber  \int_0^{+\infty}\phi(L,T) \Delta T V(L) dL  = \int_0^{+\infty}N(T) \Delta T p(L|T)V(L) dL \\
= N(T)\int_0^{+\infty}p(L|T) \Delta T V(L) dL = N(T) \Delta T V_s(T)
\end{eqnarray}
The last equality defined  $V_s(T)$, the effective average volume per cluster of given temperature in the survey. Therefore:
\begin{displaymath}
N(T) \Delta T \equiv \sum_{i}1/V_s(T_i)
\end{displaymath}
This estimator has been used by several authors (Eke et al., 1998; M98).
It is unbiased {\em provided that $V_s(T)$ is known exactly}. 
In practice, this is of course not the case, since the statistical description 
of the temperature-luminosity relation and the distribution $p(L|T)$ are not 
perfectly known, which leads to a source of {\em systematic uncertainty}. 
Consequently, a specific choice of $p(L|T)$ implies a bias in the above 
estimator. Furthermore, as fewer clusters are 
actually present in the sample, the Poisson noise increases. 
It is clear therefore, that this approach should be avoided when 
possible, unless one can be confident that the resulting uncertainties 
remain small.

A second important issue concerns errors on 
temperature measurements:  when the true underlying cluster 
distribution is very steep, positive
errors will move many more clusters upwards in temperature than  
negative errors move hotter clusters downwards, just because of 
the difference in the intrinsic abundances. This effect has been 
pointed out by Evrard (1989) for velocity dispersion measurements. 
A mean statistical correction was applied in the modeling 
by Eke et al.(1998), while Viana and Liddle 
(1999a) estimated the magnitude of this effect using a bootstrap 
resampling technique on Henry's sample. It is 
important to realize that when the errors on individual measurements are not 
identical, a mean correction is not necessarily
sufficient, because the weight $1/V_i$
varies significantly from one cluster to another. A further problem is 
that, in principle, the error on the temperature might correlate with
the apparent flux of a cluster, as fainter clusters will have lower 
signal--to--noise and will be those having the larger $1/V_i$ as well. 
In such cases, only a Monte-Carlo reproducing the exact 
conditions of the observations including the different integration times 
for the different sources, would in principle allow a complete separation 
of the various effects of temperature measurement errors. 

For the local temperature distribution function, the errors are small 
enough that the above issue is not a major 
problem.  However, due to larger temperature errors this might be a more critical point for the high--redshift sample.  
The bootstrap method used by Viana and Liddle 
(1999a) is certainly well adapted. Another possible way to solve 
this problem is to use a Bayesian approach:  we can take as a prior 
the distribution function of X-ray temperatures assuming  
$N(T) \propto T^{-\alpha}$, with $\alpha \approx 5$. 
Given that a cluster is observed with a measured 
temperature $T_0$ and assuming that the errors are log-normal distributed 
with a dispersion $\sigma$ (corresponding the observed $1 \sigma$ uncertainty), the a posteriori probability
 that the actual temperature 
of the cluster be $T$ is given by:
\begin{equation}
p(T|T_0) \propto N(T) \exp( -(\ln (T) - \ln (T_0))^2/2.\sigma^2) 
\end{equation}
this distribution is also log-normal with the same dispersion but 
with the most likely temperature $\overline{T}$ shifted compared to $T_0$:
\begin{equation}
\overline{T} = T_0 \exp( - \alpha \sigma^2) 
\end{equation}
The above formula may appear strange, because it seems to suggest that a 
systematic bias exists in the temperature measurements. This is actually 
not the case, since the observed number of clusters within the sample 
is rather constant among the different temperature bins. Only the estimated
 density is biased.

Another effect to take into account for cosmological applications (see section 5.3) when one wishes to relate the mass function to the temperature distribution function, is the dispersion in the $M-T$ 
relation. The importance of this effect  has been discussed by Eke et al (1998). They assumed an 
intrinsic dispersion of 20\%, but neglected the effect 
of the errors on the actual measurements considering them to be 
smaller.
However, as the effect goes as the square of the error, one should be 
cautious when dealing with these corrections: in practice, a
20\% error (or dispersion in the $M-T$ relation) 
begins to make a significant difference 
in the inferred abundance of clusters, while an amplitude of 10\% leads to a 
change that is essentially negligible. In their numerical simulations,  
Bryan and Norman (1998a) found an intrinsic dispersion of  
the order of 10\% in the $M-T$ relation, twice smaller than the  dispersion 
 adopted by Eke et al.(1998) ($\approx 20\%$). In order 
to understand the bias introduced by these effects (errors on temperature 
measurements and dispersion in the $M-T$ relation), the Bayesian correction 
is illuminating: as we have seen, such a correction is equivalent to 
a modification of the observed temperature, that is to say, the 
mass-temperature relation in the modeling (this is only an 
approximation because the Bayesian term depends on the actual slope 
of the N(T), which might vary with T and redshift, but this is a second 
order effect). The corrections to implement for observational errors 
are slightly different in 
nature though: errors vary from one cluster to another, and also may vary in a 
systematic way, for instance accordingly to the apparent flux. Furthermore, since  
the correcting factor varies in a nonlinear fashion, a 
mean correction is inadequate. 
Finally, it is worth noticing that errors for high redshift clusters are 
larger than for the low redshift clusters, and therefore a larger 
correction is necessary. The 
practical implication of such a change in the modelling will be discussed 
is section 5.  \\

\subsection{Temperature Distribution Function}

\begin{figure}
\resizebox{\hsize}{!}{\includegraphics{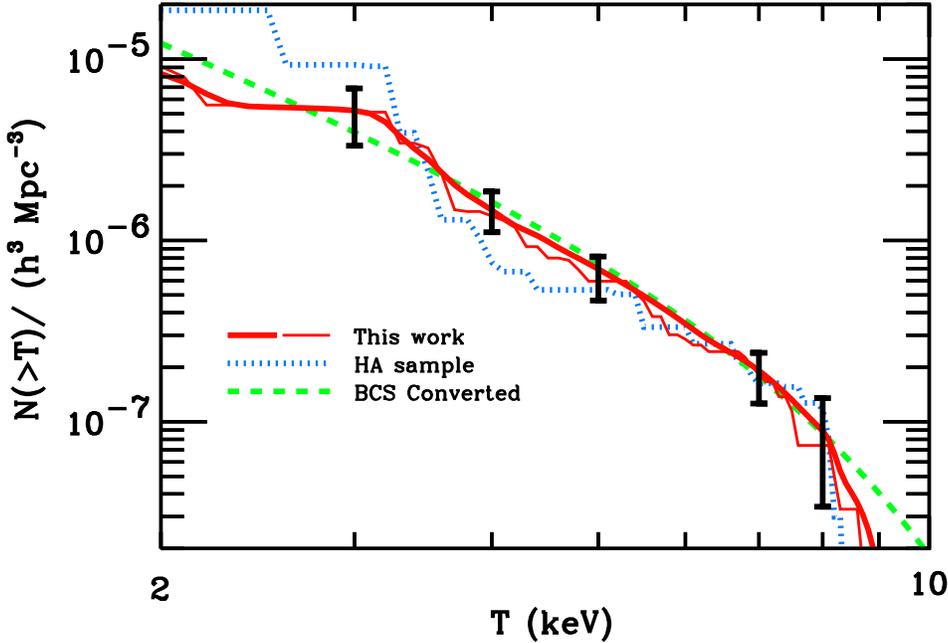}}
\caption{ \label{fig:nt1a} The integrated temperature distribution function 
inferred from our sample is given by the continuous (red) thin line. A 
smoothed version is also given (thick line). The dotted line is 
the same quantity for the original HA91 sample estimated by 
means of  Eq. \ref{eq:nesT}. 
The dashed (green) line is derived from the BCS luminosity function.
The error bars represents the 68\% interval for the distribution of the estimator from the bootstrap.}
\end{figure}

\begin{figure}
\resizebox{\hsize}{!}{\includegraphics{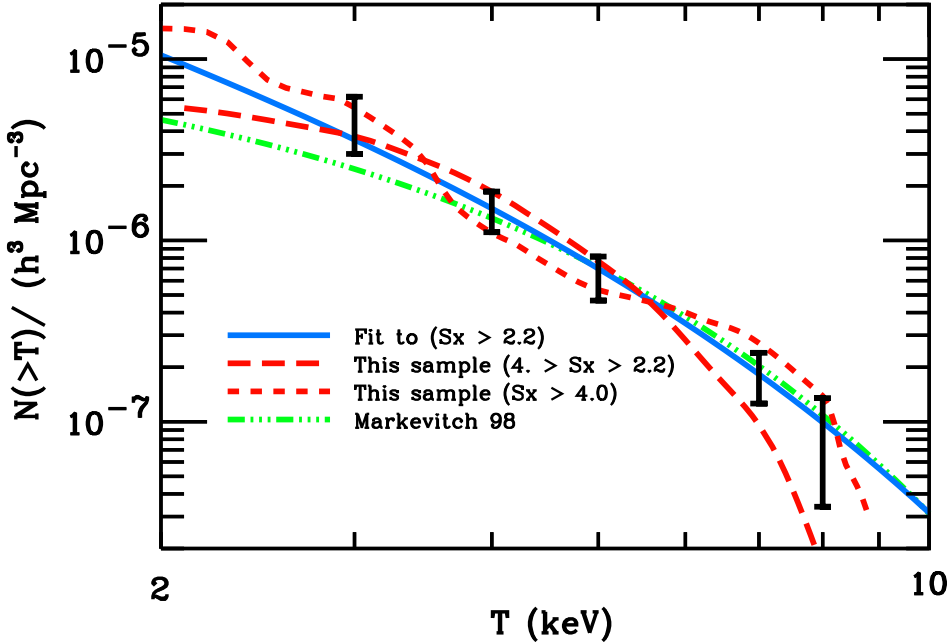}}
\caption{ \label{fig:nt1b} A fit to our temperature distribution function 
is given by the continuous (red) thick line. The two thick dashed lines 
correspond to two  sub--samples: the short dashed line corresponds to the bright one ($2.2 10^{-11} > s_x > 4. 10^{-11}$ erg/s/cm$^2$), the long dashed corresponds to the faint sub-sample ($s_x > 2.2 10^{-11}$ erg/s/cm$^2$). The long-dashed-3-dotted 
(green) line is the temperature distribution function from Markevitch (1998).}
\end{figure}

We compute the integrated temperature distribution from 
our sample following the method described in the previous section, i.e. using the estimator:
\begin{displaymath}
\nonumber
N(>T) \equiv \sum_{\overline{T}_i > T}1/V(\overline{T}_i)
\end{displaymath}
The resulting temperature distribution function is plotted as the 
continuous thin line in Figure 
\ref{fig:nt1a}. A smoothed version is also given. We have also checked that using
temperature when the central emission is removed makes negligible difference.
For comparison, 
we have also estimated the temperature distribution 
function from the HA91 sample, using available updated temperature 
measurements and the standard estimator (Eq. \ref{eq:nesT}) without 
the Bayesian correction factor. This temperature distribution is shown as 
the dotted line in Fig. \ref{fig:nt1a}. As one can see, the difference is quite noticeable, in 
particular around 4 keV, where it is roughly a factor of two. We also show the temperature distribution inferred from the BCS luminosity function 
(Ebeling et al, 1997) assuming the $L_x-T_x$ relation 
as given by OB97. A convenient fit of our temperature distribution function is:
\begin{equation}
N(>T) = 1.42 10^{-6}h^3 {\rm Mpc}^{-3} \left(T/4. {\rm keV}\right)^{-1.6}\exp \left( -(T-4. {\rm keV})/2.6 \right)
\label{eq:ourNT}
\end{equation}
which is plotted in Figure \ref{fig:nt1b}. 
We also show the fit to the temperature distribution function derived by M98. 
As one can see, we find
a larger abundance of cool clusters ($T \leq 4$ keV) than 
M98 while we have a nearly 
perfect agreement with the temperature distribution function 
inferred from the BCS luminosity function.\\

The difference between our temperature distribution function and 
previous estimations is striking. The 
difference with M98 can be easily understood -- his sample being  
restricted to the redshift range $z > 0.04$, a significant fraction of 
faint (and cool) clusters are missing. He has corrected for this 
incompleteness by a weighting scheme based on an estimated dispersion in the 
$L_x-T_x$ relation and following the method discussed in section 4.1. However, in 
order to properly evaluate such a correction, one needs an accurate estimate 
of the dispersion in the $L_x-T_x$ relation, which can only be obtained 
from a flux--limited sample. The difference with the original HA91 sample 
is more difficult to understand. One may argue that the  
difference comes essentially from the band selection ($2-10$ keV ), however as we have already mentioned in section 2, comparison with the 
E-HA sample suggests another origin. 
We have therefore divided our sample into two equal 
sub-samples, corresponding to the brightest and the faintest 
clusters (in apparent flux). Again, since both samples are X-ray 
selected (in the same band), they should be statistically equivalent. 
The corresponding temperature distribution functions are shown in Figure 
\ref{fig:nt1b}. The bright sub-sample leads to a temperature distribution 
function close to the one based on the HA91 sample. This is not surprising
on the high temperature end, since the clusters are almost the same in 
these two samples. The fact that the temperature distribution functions 
are almost the same for the cooler clusters as well, indicates that the 
original HA91 sample does not suffer from any significant bias and that 
the effects of the chosen band are relatively small. On the other hand, the
difference between the faint and the bright samples is quite significant, 
and the relative abundances of hot and cool clusters is noticeably different. 
As the difference is marginally 1 $\sigma$, these two 
samples could be considered as representing two realizations which are 
slightly in the tail of the distribution, without being really suspicious. 
Nevertheless, such a difference have some implications in practice as we will see in the next section. The fact that our temperature distribution function is in good agreement 
with that inferred from the BCS luminosity function is rather encouraging 
and confirms the absence of any systematics that would undermine our analysis.
 
Differences such those we have found have significant  
consequences when constraining the spectrum of the 
fluctuations from the temperature distribution function: a sample in which 
the abundance of cool clusters is underestimated leads to a too flat 
spectrum and the amplitude of fluctuations at cluster scales will be also underestimated.  All these effects enter as a 
source of systematic uncertainty in the estimation of $\Omega_0$.\\

\subsection{ Distribution function of the estimator and Confidence intervals }

\begin{figure}
\resizebox{\hsize}{!}{\includegraphics{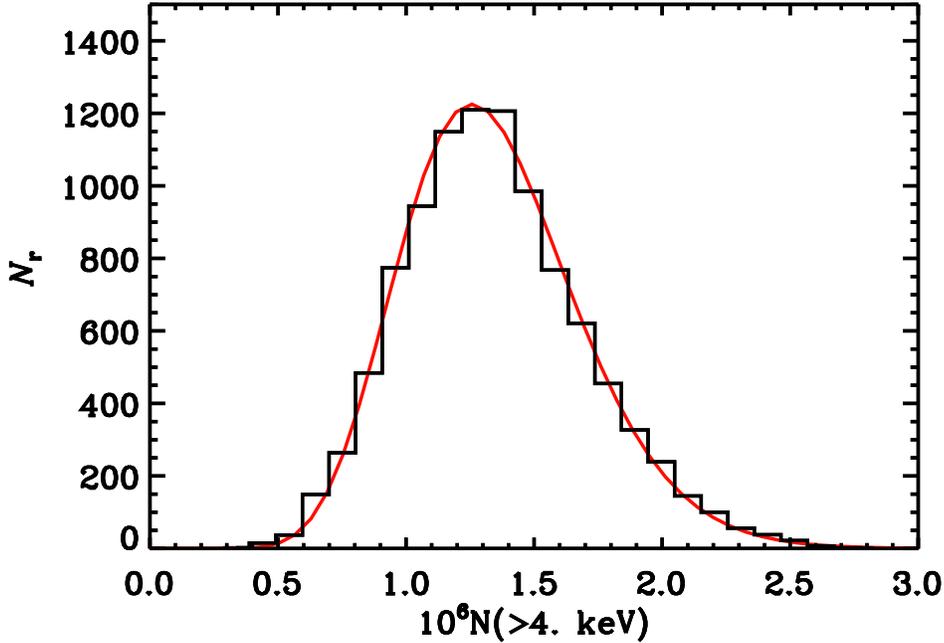}}
\caption{ \label{fig:distri} The finite size of the sample 
implies that our estimate of the abundance of clusters has some uncertainty.
The distribution of values that our estimator can take, given the sample we 
have, is estimated by a Bayesian bootstrap.
Here we illustrate the histogram of the values  taken 
by our estimator  of the 
abundance of clusters  at present epoch with temperature above 4 keV . A convenient analytical fit of 
this distribution is also shown: a $\chi^2$ distribution with 6 degrees of freedom.}
\end{figure}

It is useful to have a statistical description as comprehensive as 
possible in order to derive strong constraints on models. The  number 
$\cal N$ of clusters in a given temperature range can be regarded as a 
Poisson realization of mean $\int_0^{+\infty}$ $\phi(L)$ V(L) $\Delta$ L. 
However, the understanding of this quantity requires good information 
on the $L_x-T_x$ relation: its shape, its  normalization and the distribution function around it. In order to determine the expected distribution of 
our estimator for $N(T)$, we use a Bayesian bootstrap technique that 
allows us to avoid the use of a specific $L_x-T_x$ relation, getting rid of some possible bias thereby introduced. We therefore 
assume that the distribution of the estimator $ \tilde{N}(>T)$ for a given 
density $ N(>T)$ is of the following form:
\begin{equation}
D(\tilde{N})d\tilde{N} =  d(\lambda)d\lambda
\label{eq:distri}
\end{equation}
with $\lambda = \tilde{N} / N(>T)$. We then use a Bayesian bootstrap to 
reconstruct the distribution function $d$: $10^{4}$ fake samples are built 
from the original sample, each having the same mean number of clusters, but 
with a given dispersion to account for Poisson noise.
This procedure was used to find the distribution function of the 
estimator of $N(>T)$, as well as of the estimator for $N(>T_1)-N(>T_2)$. 
The latter quantity is easier to handle with in a statistical analysis,
as different temperatures can be chosen in such a way that the 
measurements are independent 
quantities (typically $N(>T_1)-N(>T_2)$, $N(>T_2)-N(>T_3)$,...). 
We found that a $\chi^2$ distribution function fits quite well the 
distribution function of the 
estimator, provided that the number of degrees of freedom is left as a free 
parameter (and is not necessary an integer). An example is given in 
Figure \ref{fig:distri}. As one 
can see, the $\chi^2$ distribution provides an adequate fit to the overall 
distribution.  From this we can infer confidence intervals on the density. 
The error bars given in Figure \ref{fig:nt1a} reflect the 68\% 
range  of the value of the estimator of the  density of clusters. 

\section{ Cosmological Applications}

\subsection{Mass-temperature relation}

In order to relate the observable properties of clusters to their 
mass, physical correlations are necessary.  As discussed in the 
introduction, the temperature of the X-ray gas is probably the most 
convenient cluster mass estimator. Kaiser (1986) developed the scaling 
arguments which are essential to this approach (they were already used 
in 1980 in the pioneering by Perrenod, 1980).  
Some of the scaling arguments are exact, and they are remarkably powerful.
In an Einstein--de Sitter universe with initial fluctuations described by a power--law spectrum, the only 
process acting is gravity and therefore for 
the physics of the gas, scaling arguments will hold as long as no other 
mechanism plays an important role (like cooling or bulk 
heating). In such a situation, the only possible scale of the problem is 
associated with the non-linear scale, even in the presence of 
shock heating.  This guarantees that the 
$T-M$ relation for $M_*$ clusters follows an exact scaling law (for a given 
spectrum):
\begin{displaymath}
T_* \propto M_*/R* \propto M_*^{2/3}(1+z)
\end{displaymath}
It is reasonable to expect that such a relation will also hold between 
different masses forming at a given redshift and that is nearly 
independent of the power spectrum. However, this remains an approximation 
as the geometrical aspect of the peaks corresponding to different masses at 
a given time may differ (due to the difference in the height of 
the peaks). The same remark applies for two different spectra. 
Further checks are therefore needed. 
Analytical modeling of clusters adopting hydrostatic 
equilibrium is often completed by the isothermal assumption. 
However, such a mass estimator 
can be misleading, even when temperature profiles are taken into account 
(Balland \& Blanchard, 1997). A further problem is the fact that a 
significant part of the pressure support of the gas could come 
from turbulent motions in the gas (Bryan and Norman, 1998b). Furthermore, 
the scaling laws are not expected to hold exactly in open models, significant departures may actually exist in 
this case (Voit \& Donahue, 1998). It seems therefore safer to rely on 
numerical simulations in order to establish the mass-temperature relation. 
There are several ways to define the mass of clusters, which are not 
actually objects with well defined and sharp boundaries.  
Considerable progress has been made in recent years thanks to Navarro, Frenk and White (1996), who showed that clusters are well fitted by 
a universal profile involving few parameters (which is not what would be 
expected based on the isothermal hydrostatic $\beta$-model). Still, 
various definitions of the mass are used, resulting in some confusion 
when one wishes to make detailed comparisons. The mass of a cluster could be defined as the mass within the so-called virial radius corresponding to a spherical region with a contrast 
density $\delta=178$ (in the Einstein-de Sitter case). The mass can also be defined by $\delta=200$, $\delta=500$, or by the mass within a fixed physical radius such as the Abell radius or its equivalent in comoving coordinates. As 
most of the numerical checks of the mass function have been performed at 
the virial radius, it is certainly safer to estimate the mass at this 
very radius. The validity of the scaling 
models has been impressively demonstrated (Evrard 1989), and no noticeable difference 
has been detected among different spectra, masses and cosmological backgrounds. 
Furthermore, an intensive comparison of the results from various codes 
has verified that different numerical techniques do not lead to significant 
differences in inferred properties (Frenk et al., 1999). This is 
especially true for the $M-T$ relation. Even the inclusion of a
significant energy injection does not introduce significant changes in 
these relations (Metzler and Evrard, 1998). The recent numerical work of 
Bryan and Norman (1998a) is certainly
the most advanced in this area, and the scaling relations they obtained are 
impressive (see their figure 4) and convincingly independent of the 
underlying cosmological model.  Their result can be summarized by the 
fact that
  the following mass-temperature relation:
\begin{displaymath}
T_v = 3.8 {\textrm keV}  
  M_{15}^{2/3}(\Omega_0\Delta/178)^{1/3}(1+z)  
\end{displaymath}
(given for $h =0.5$) reproduces well the results of their numerical simulations ($M_{15}
\equiv M/10^{15}$ solar masses). The statistical  
dispersion they found around this relation is only 10\%. The slope 
$2/3$ might not be guaranteed to better than 15\%.  However, 
the normalization of the above relation in the various numerical 
simulations can differ (for instance, Evrard et al., 1996 found a value 
20\% higher than Bryan \& Norman, 1998a, while the mean value obtained 
by Frenk et al., 1999, is only 5\% higher with a 5\% dispersion among 
the various simulations). The existence of turbulence on 
small scales (Bryan and Norman; 1998b) indicates that one should be 
cautious on this issue, as the modeling of turbulence is often 
limited by the resolution of the simulation. Finally, it is 
important to notice that such masses are 70\%  higher than what 
would be inferred from the isothermal hydrostatic equation (Roussel et al., 2000) and more than twice the mass inferred  recently from 
 decreasing temperature 
profiles as considered by M98, or more recently by 
Nevalainen et al. (1999). In our modeling, we 
have adopted the above normalization, corrected for the effect of 
introducing a 10\% dispersion in the $M-T$ relation:
\begin{displaymath}
T_v = 4.0  {\textrm keV}  M_{15}^{2/3}(\Omega_0\Delta/178)^{1/3}(1+Z)  
\end{displaymath}
($h =0.5$). We will examine the consequences of any change in this relation.

\subsection{Constraining the models}

\begin{figure}
\resizebox{\hsize}{!}{\includegraphics{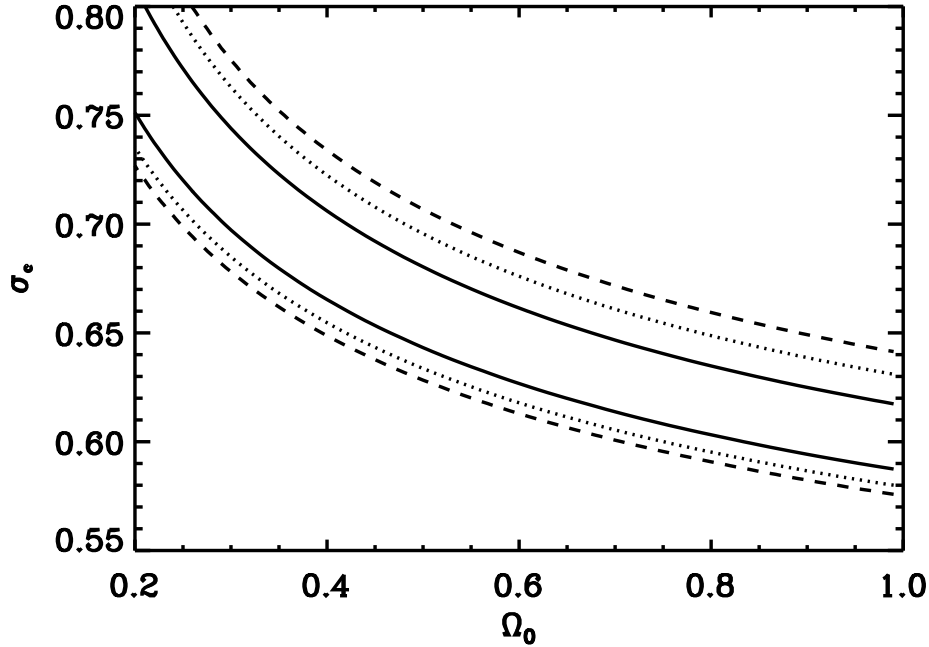}}
\caption{ \label{fig:sigom} Contours giving the allowed range of values for 
the amplitude of matter density fluctuations on cluster scales. The 
continuous line is the 68\% contour, the dotted line is the 90\% contour, 
and the dashed line is the contour at 95\%.}
\end{figure}

\begin{figure}
\resizebox{\hsize}{!}{\includegraphics{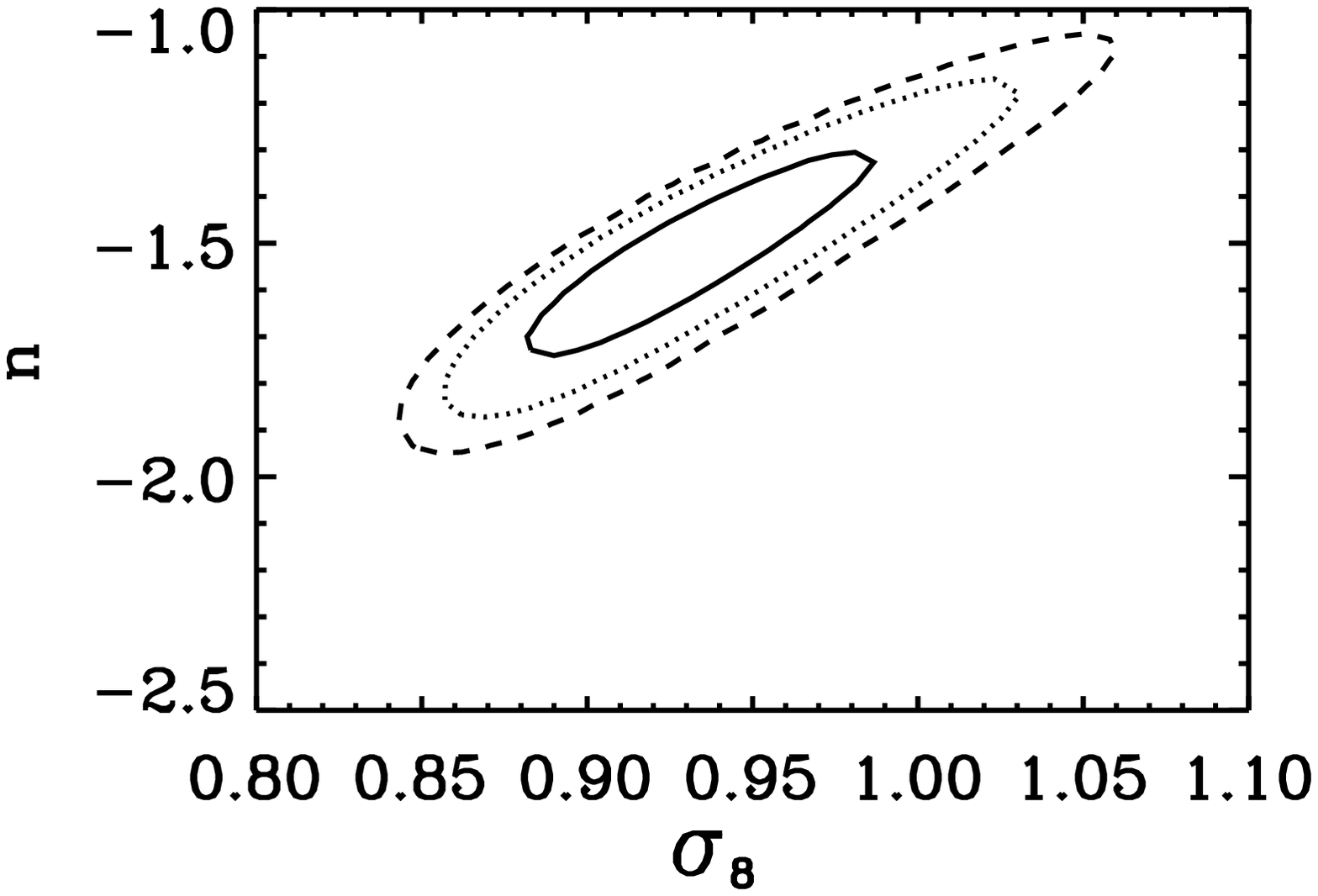}\includegraphics{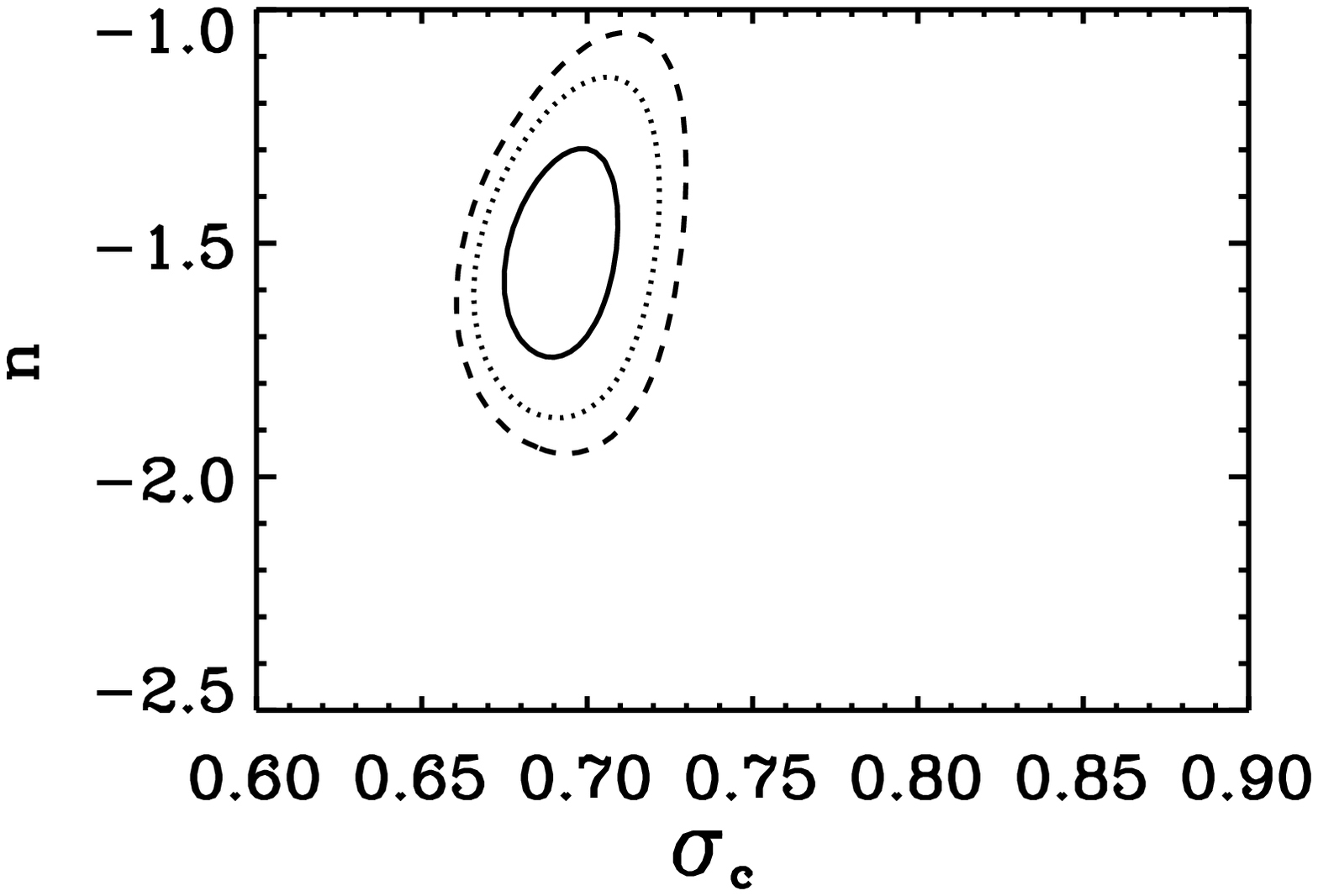}}
\caption{ \label{fig:norm} Contours giving the range of allowed values for 
the index $n$ of the primordial matter fluctuation spectrum on cluster scales
for a Gaussian distribution function versus the normalization expressed in term of $\sigma_8$ (left side) and in term of $\sigma_c$ for a density parameter  
$\Omega_0 = 0.4$. The continuous lines are 
the 68\% contours, the dotted lines are the 90\% contours, and the 
dashed lines 
are the contours at 95\%. }
\end{figure}

The local temperature distribution function can be used to constrain the 
properties of the matter density fluctuations.  Although this 
can also be applied to a non-Gaussian scenario (OB97; Robinson et al., 1999), 
we will restrict ourselves to Gaussian fluctuation models only. 
This consists basically in finding the 
best parameters and confidence intervals from a fit to the 
temperature distribution function.  However, there is a degeneracy 
between different cosmological models: equally satisfying fits can be 
obtained in either low- or high--density  
models (OB92). Nevertheless, the density fluctuation amplitude can 
be usefully constrained for a given model. It is generally assumed 
that cluster abundances lead to a determination of the fluctuation
amplitude at $8h^{-1}$Mpc and therefore the normalization is communly 
expressed in term of $\sigma_8$. This is not completely correct. 
In an $\Omega_0 = 1$ universe, the temperature 
distribution function of clusters does actually provide 
the normalization of the fluctuations on this scale. This comes from 
the fact that the mass of a $8h^{-1}$Mpc sphere collapsing today will 
lead to a 4.5 keV cluster, which is a typical temperature of a moderately 
rich cluster for which X-ray 
samples provide a reasonably well established abundance 
(at least this what the authors now believe!). In contrast, in a 
low--density universe ($\Omega_0 \sim 0.3$), a $8h^{-1}$Mpc sphere will 
produce a 2 keV cluster, for which we actually do not yet have reliable 
abundance estimates. This is not just a semantic difference, 
since for a low--density universe ($\Omega_0 \sim 0.2$), the  amplitude of  
$\sigma_8$ could  
differ by up 50\% (OB97).  Following Blanchard \&  Bartlett (1998), 
we will use the following definition:
\begin{displaymath}
\sigma_c = \sigma(\Omega_0^{-1/3}8h^{-1}\textrm{Mpc})
\end{displaymath}
and give constraints in terms of $\sigma_c$.  This has the additional 
advantage 
of being independent of the power spectrum: as $\sigma_8$ and the power index $n$ are 
correlated, the uncertainty on $\sigma_8$ is larger than on $\sigma_c$. In order to illustrate this point we have 
plotted both quantities in Fig. \ref{fig:norm}.

To find parameter constraints, we find the best model according to 
the following likelihood function:
\begin{equation}
{\cal L} = \prod p\left(N(>T_i,z_i)-N(>T_{i+1},z_i)| N_{\{\Omega_0,n,\sigma_c\}}(>T_i,z_i)-N_{\{\Omega_0,n,\sigma_c\}}(>T_{i+1},z_i)\right)
\end{equation}
in which $N$ is the measured abundance and $N_{\{\Omega_0,n,\sigma_c\}}$ is the value given 
by the model, which depends on the density parameter $\Omega_0$, 
the local slope of the power 
spectrum $n$ and its amplitude specified by $\sigma_c$.
The probability distribution function $p$ cannot be inferred from 
first principles and is therefore estimated with our bootstrap procedure. 
The maximum value of the likelihood has been normalized to unity, and 
confidence intervals on the parameters were estimated from the  
contours at ${\cal 
L }= 0.6, 0.26, 0.13$ corresponding respectively to the 68\%, 
90\% and 95\% contours (strictly speaking this holds only for normal distributions).  The constraint on the shape of the spectrum 
$n$ is rather poor.  We find a broad range of possible values. 
The only obvious result is that models with $\Omega_0 = 1$ and  
$n\sim -1$, representing the standard Cold Dark Matter spectrum,
are excluded, as previously found (Blanchard \&  Silk, 1991; 
Bartlett \& Silk, 1993). The 
effect of statistical uncertainties on the amplitude of matter 
fluctuations on cluster scales is quite small, and therefore only 
a restricted range of values is allowed for a given $\Omega_0$. 
The dependence on $\Omega_0$ is weak :
\begin{eqnarray}
\nonumber \sigma_c \approx  & \: 0.6~^{+0.035}_{-0.025}  \:\Omega_0^{-0.14} \: (95\%)& \textrm{open case}\\
\sigma_c \approx & \:0.6~^{+0.035}_{-0.025}  \:\Omega_0^{-0.18} \: (95\%) & \:\textrm{flat case}
\end{eqnarray}
(uncertainties are statistical and are given for $\Omega_0 = 1$). 
Different values have been published for the amplitude obtained from the 
cluster normalization. These differences are mostly due to different cluster abundances used. Our 
normalization is slightly higher than what is often derived from abundances 
of X-ray clusters (but in better agreement with some normalizations based on 
optical surveys, e.g., Girardi et al, 1998). This is due to 
the fact that our number density of clusters at 4 keV is higher than previous estimations and that we used the Bryan and Norman (1998) normalization of 
the $M-T$ relation which is slightly lower than that of Evrard et al. (1996).  Finally, the differences with previous works concerning 
the dependence on $\Omega_0$ come from the fact that we 
refer to $\sigma_c$, which is the normalization on the true cluster scale rather than to $\sigma_8$, which depends on the assumed 
spectrum (for $n =-1.5$ we find $\sigma_8 = 0.96$ in an open $\Omega_0 =0.3$ 
universe).  

It is interesting to try to evaluate the magnitude of possible 
systematic uncertainties on $\sigma_c$. As we have 
argued, the local cluster abundance now seems well established, 
so that we do not expect this to be a source of systematic error
larger than the statistical uncertainty already quoted. The remaining 
dominant uncertainty comes from the use of the Press and Schechter mass 
function. Different results exist in the literature concerning
its accuracy to describe the actual mass function 
(Lacey \& Cole, 1994). 
Numerical simulations mostly suggest that the mass function is well described 
by the Press and Schechter formula: for instance,    
Borgani et al. (1998) have shown  that it can be safely used for cluster   modeling in arbitrary cosmology.

\begin{figure}
\resizebox{\hsize}{!}{\includegraphics{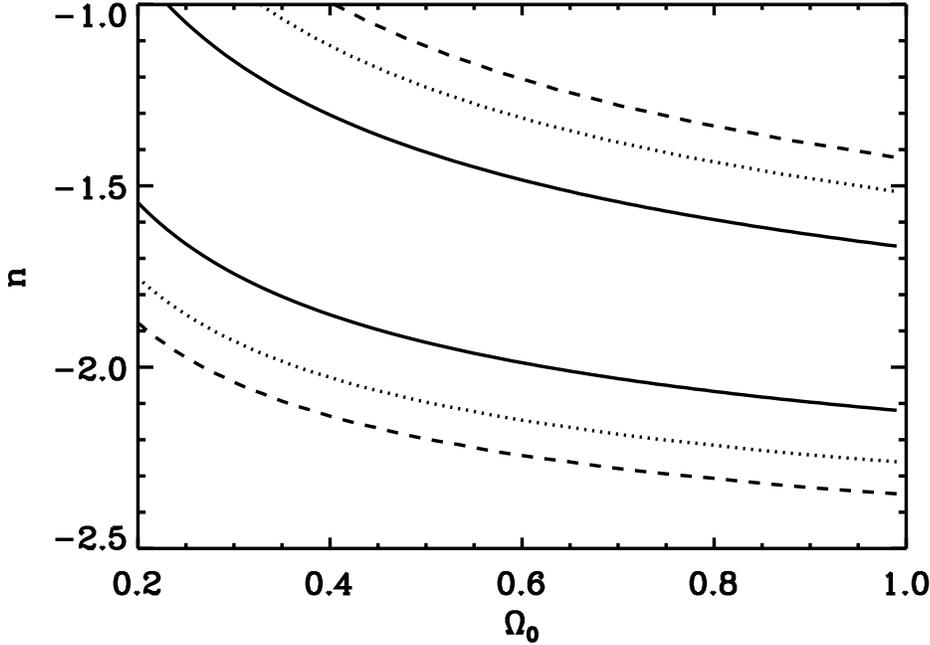}}
\caption{ \label{fig:nom} Contours giving the range of allowed values for 
the index $n$ of the primordial matter fluctuation spectrum on cluster scales
for a Gaussian distribution function. The continuous line is 
the 68\% contour, the dotted line is the 90\% contour, and the dashed line 
is the contour at 95\%. }
\end{figure}

\subsection{Estimating $\Omega_0$}

The determination of the mean density of the universe is obviously 
a very important goal of observational Cosmology. Although it is
fashionable to claim that current data favor a low value, it seems to
us that the observational situation has not improved that much since the 
1980's. Indeed, bulk flows on large scales lead to somewhat contradictory 
results and interestingly enough, 
fluctuations in the CMB seem to favor flat models and to exclude open 
ones (Lineweaver et al., 1997; Lineweaver and Barbosa, 1998; Le Dour et al., 2000). 
Perhaps the most stringent constraint comes from the baryon fraction 
observed in clusters (White et al, 1993). Still, a firm upper limit on 
$\Omega_0$ is lacking, because of the uncertainties in primordial 
nucleosynthesis and cluster mass estimates (notice that 
virial masses estimated from the scaling models with the normalization we adopt are
nearly twice as large as those based on hydrostatic estimations!).\\

The evolution of clusters is known to be a powerful probe of the mean
density of the universe (Perrenod, 1980; Peebles et al. 1989).  Evolution
of the cluster abundance with redshift offers a new pass to $\Omega_0$ (OB92, OB97) which is 
insensitive to the cosmological constant, and above everything else
is really a global probe rather than a local one, as it is the case with the classical $M/L$ argument. It has been shown that the EMSS 
distribution can be modeled equally well within low- and high--density 
universes, provided that some evolution in the 
$Lx-Tx$ relation is adopted (OB97). The required evolution depends on 
$\Omega_0$ as follows:
\begin{equation}
 L_x \propto T_x^3 (1+z)^B
\end{equation}
with
\begin{equation}
 B \sim 4 \Omega_0 - 3.
\end{equation}

It is important to ask whether this modeling based on limited 
observational data is robust or not. Since this work, the 
observational situation has been considerably improved, 
thanks to the ROSAT satellite. 
The number counts for X-ray clusters are now available and fall well within the range of the predictions given by OB97. 
The ROSAT X-ray cluster luminosity function was derived from the 
BCS sample (Ebeling et al. 1997) and recently confirmed by De Grandi et al. 
(1999a). This observed luminosity 
function matches extremely well the one predicted by OB97. Finally, the amplitude of the fluctuations which was 
normalized to the original HA91 temperature distribution function, is in good agreement with what has been obtained in the present work. This 
considerably reinforces the strengt of OB97 modeling. Reichart at al. (1999) have recently re-analyzed the EMSS sample and have reached similar conclusions. Furthermore, similar value of $B$ has been obtained 
by Borgani et al. (1998) from their analysis of the RDCS survey. 
The determination of $\Omega_0$ from 
the evolution of the $L_x-T_x$ relation seems therefore possible, 
and we expect it to lead to reliable estimates. Thanks to the availability 
of ASCA temperature measurements for a significant number of high--redshift 
clusters, it has been shown that this relation undergoes at most only
a mild evolution (in the positive direction, i.e., high--redshift clusters 
are more luminous than local ones with the same luminosity; Sadat 
et al., 1998). Using the relation between B and $\Omega_0$ (eq. 9),
Sadat et al. (1998) concluded in favor of a high value of the density parameter ($\Omega_0 \sim 0.8$) with a quoted uncertainty of $0.2$ 
(this uncertainty only reflects the uncertainty due to the Poisson noise 
in the EMSS distribution, and does not  
take into account the uncertainty in the overall modeling). \\

Thanks to the availability of the first X-ray selected sample of 
high--redshift clusters with measured temperatures (H97), it is now  
possible to constrain the density of the universe using directly the evolution of the X-ray cluster abundance to an intermediate 
redshift. A value of $\Omega_0 = 0.5$ was obtained (H97) (this value has been abusively referred 
to as a low $\Omega_0$.). Eke et al. (1998) who studied in detail the 
various systematic uncertainties have found a similar  $\Omega_0$. A higher value of $\Omega_0$ has been derived by Viana and Liddle (1999a) who discussed the importance of a proper correction from the uncertainties introduced by temperature measurement errors. Since these authors used a 
lower local abundance than the one we have inferred from the present work, it is interesting to re-address 
this question. Our primary goal here is to analyze the consequences 
of the higher cluster abundance at zero redshift that we derive. 
We leave to a future work a more complete investigation of the 
high--redshift sample. 

\begin{figure}
\resizebox{\hsize}{!}{\includegraphics{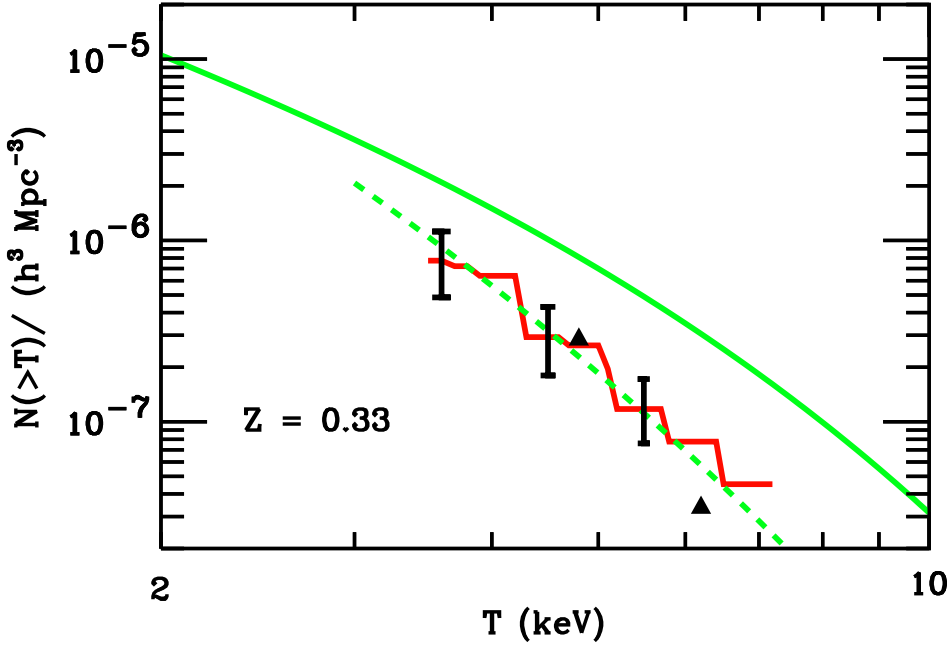}}
\caption{ \label{fig:nt2} The integrated temperature distribution function 
at $z = 0.33$ inferred from Henry's sample is given by the continuous thick (red) 
line (corresponding to the case $\Omega_0 = 0.5$).  The two triangles are 
the estimation from Viana \& Liddle (1999a, 1999b) with the original H97 
sample.
The line is the best fitting model. The continuous grey (green) line is the fit to our  local temperature distribution function. The dashed grey (green) line is our best model at $z =0.33$.
}
\end{figure}

We have estimated the temperature distribution function
of the distant cluster sample as given by H97, and estimated the 
suitable volume correction for the various cosmological models. Temperatures 
were corrected by the Bayesian term. The resulting temperature distribution 
function is given in Figure \ref{fig:nt2}. 
For comparison, we have also plotted the temperature distribution function 
inferred  from the local luminosity function. One may worry that 
our Bayesian correction is adequate. The bootstrap approach followed by 
Viana and Liddle (1999a) is certainly well adapted to the treatment of 
measurement errors.  We have therefore compared their inferred abundances 
to ours and found very good agreement (using the 10 original clusters). 
This confirms that our analytic method provides a correction term which is 
as good as the bootstrap resampling technique. Our inferred temperature 
distribution is shown as the thick continuous line
in Figure \ref{fig:nt2}.

\begin{figure*}
\resizebox{\hsize}{!}{\includegraphics{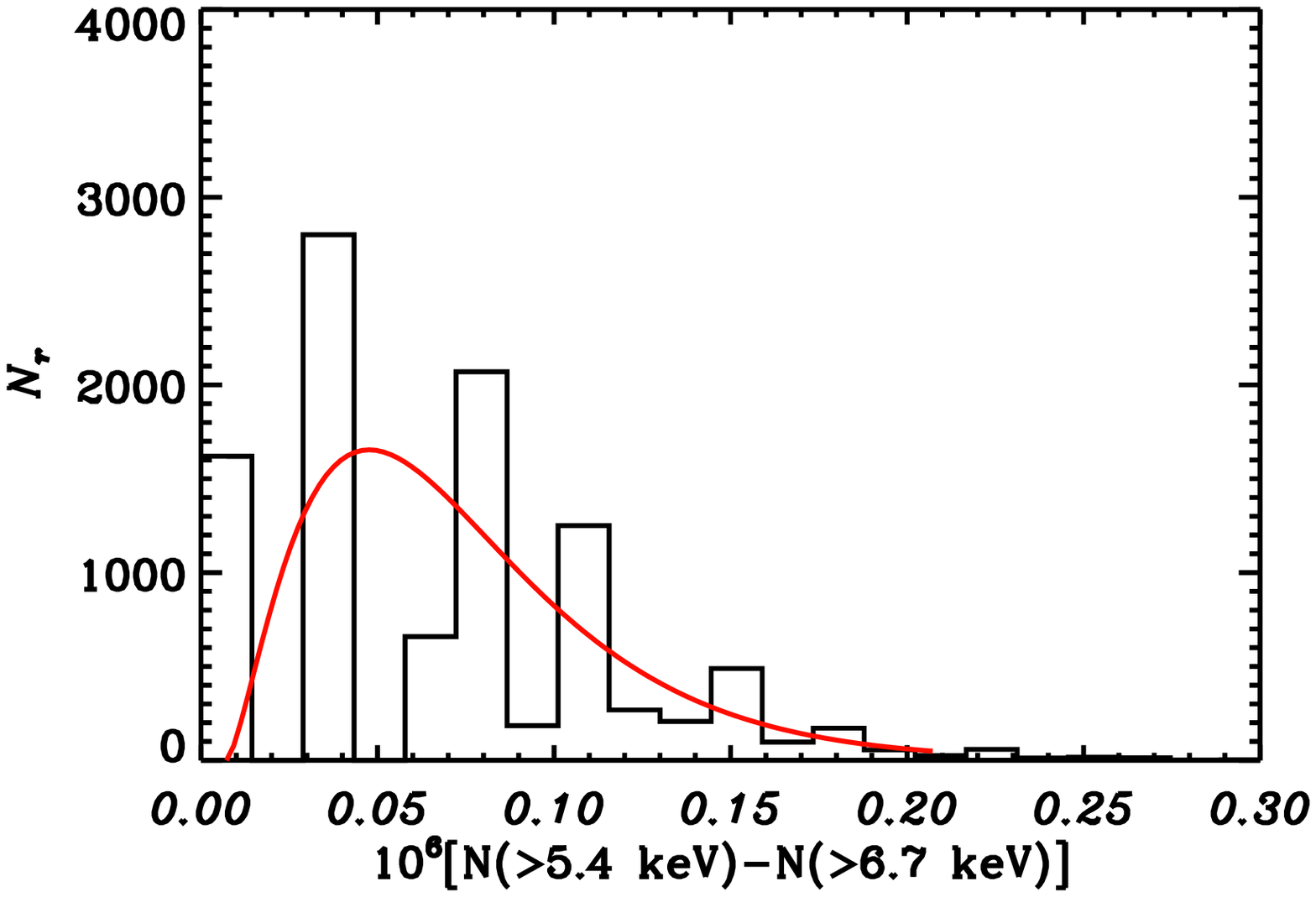}
\includegraphics{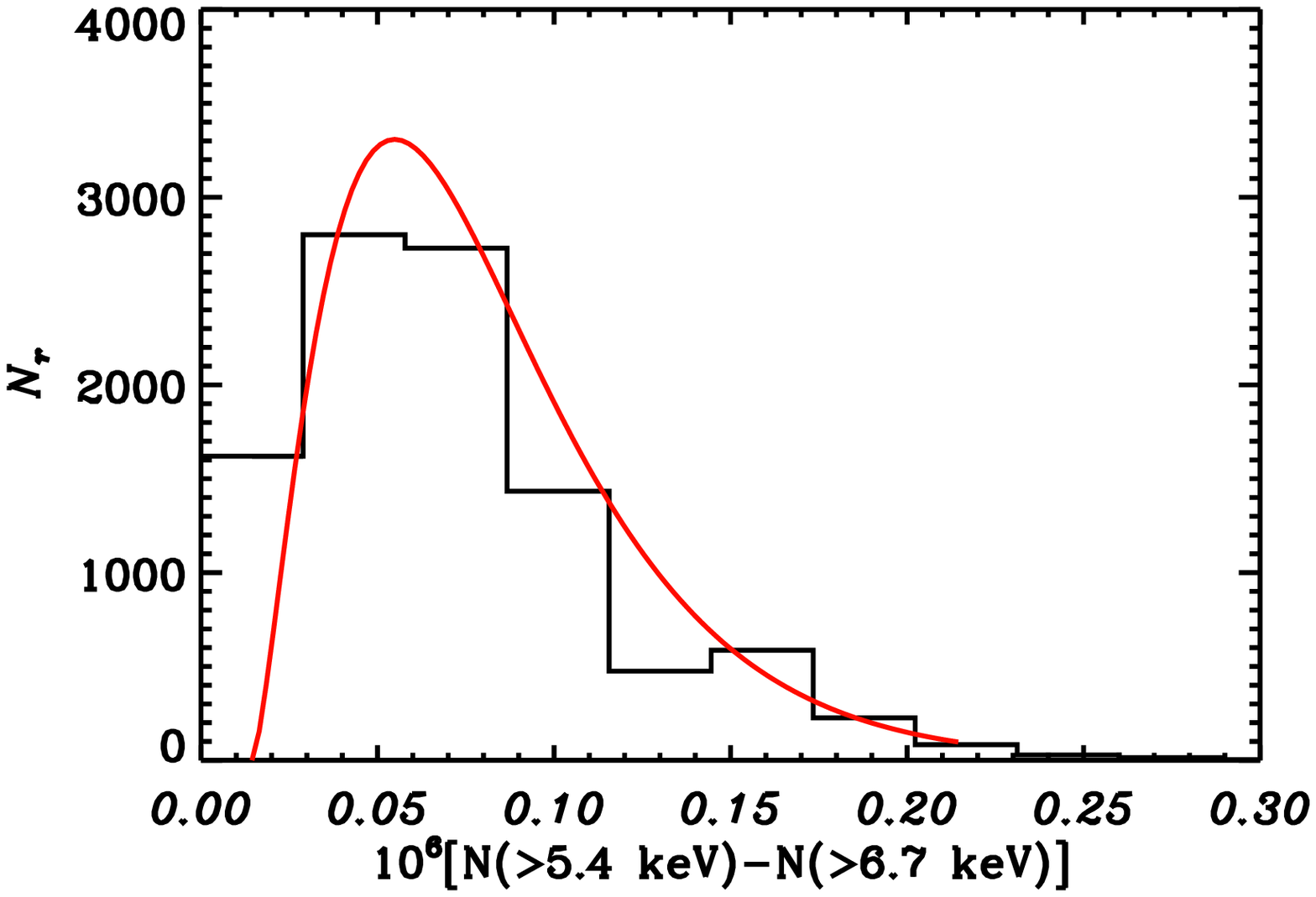}}
\hspace*{2.8cm}a) \hspace*{6.4cm} b)
\caption{  \label{fig:distriz03} Figure a) represents the histogram 
of the values taken by our estimator from $10^4$ bootstrap resampled samples (as in figure \ref{fig:distri}). The spikes are real 
and reflect the fact that the abundance estimation is dominated by a single 
cluster. In this case, our analytic expression for the distribution function 
does not work very well (continuous line : 
$\chi^2$ distribution with 1.5 dof). b) represents the distribution function of 
the estimator with larger bins; in this case the fit is acceptable.}
\end{figure*}

We use the following likelihood function to infer the best-fit 
value of $\Omega_0$:
\begin{eqnarray}
\nonumber \lefteqn{
{\cal L} = \prod_{i}  p\left(N(>T_i,z_i)-N(>T_{i+1},z_i)| N_{\{\Omega_0,n,\sigma_c\}}(>T_i,z_i)-N_{\{\Omega_0,n,\sigma_c\}}(>T_{i+1},z_i)\right)
}\\
& & \prod_{j}  p\left(N(>T_j,z_j)-N(>T_{j+1},z_j)| N_{\{\Omega_0,n,\sigma_c\}}(>T_j,z_j)-N_{\{\Omega_0,n,\sigma_c\}}(>T_{j+1},z_j)\right)
\end{eqnarray} 
with $T_i=\{3., 4., 5.4, 6.6, 8., +\infty, z_i = 0.05 \}$ and 
$T_j=\{3.6, 4.4, 5.5, 6.7, 8., +\infty, z_j = 0.33\}$. 
The distribution function of 
the measured abundances was estimated by a Bayesian bootstrap, as 
discussed in section 4.3. Because of the small number of clusters in 
the sample, the inferred abundance looks very spiky (see Figure 
\ref{fig:distriz03}a), which simply means that one cluster 
essentially dominates the statistic. In such a case, our fits are of course 
never as good as in the case where the number of clusters is larger, 
but they are reasonably acceptable when the binning in abundance is 
enlarged (see Figure \ref{fig:distriz03}b).

\begin{figure}
\resizebox{\hsize}{!}{\includegraphics{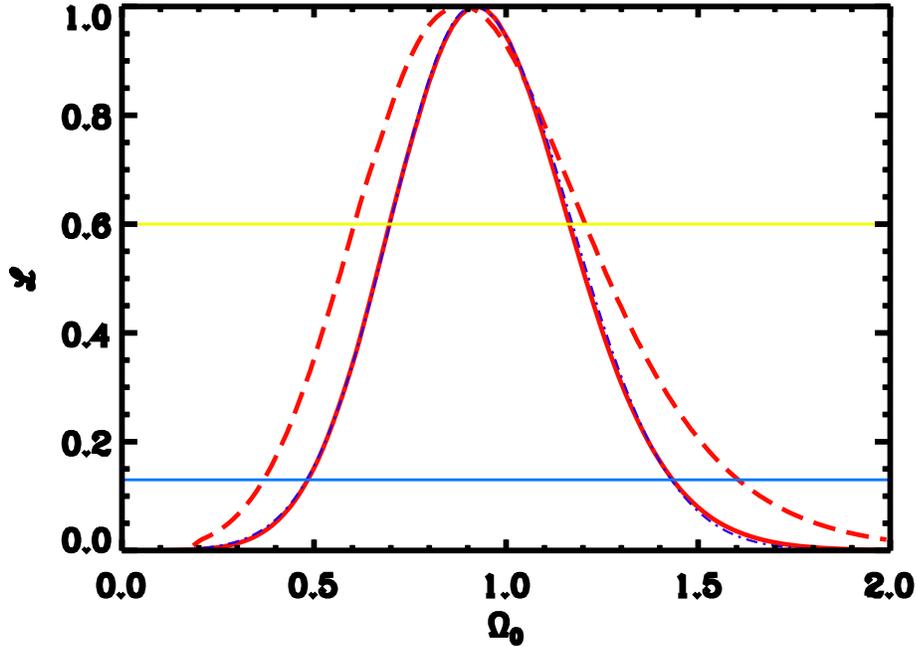}}
\caption{ \label{fig:Lom} The likelihood function provided by the comparison 
of the abundances of clusters at $z = 0.05$ and $z = 0.33$. The continuous 
line corresponds to the open case, the dashed line corresponds to the flat 
case. The two horizontal line gives the 1$\sigma$ and 2$\sigma$ confidence 
interval on one parameter for a gaussian distribution. The dotted-dashed 
line is a fit by a two-tailled gaussian. }
\end{figure}

 The 
likelihood function normalized to unity, is given in Figure \ref{fig:Lom}. 
The most likely values  of $\Omega_0$ are 0.92 (open case) and 0.865 (flat case). 
The shape of the likelihood function 
is very well fitted by a Gaussian, even if the probability functions we have 
used are significantly non-Gaussian. We can therefore make direct 
use of this function to give the confidence intervals at the $1 \sigma$ level. 
Our best fitting values are :
\begin{eqnarray}
\nonumber \Omega_0 = & 0.92^{+0.255}_{-0.215}   &  \mbox{  {\textrm (open case)}}\\
 \Omega_0 = & 0.865^{+0.35}_{-0.245}     & \mbox{ {\textrm (flat case)}}
\end{eqnarray}
This is very consistent with Sadat et al. (1998) value
from a completely independent analysis. The constraint we have
obtained is quite severe: an open model with $\Omega_0 < 0.49$ is ruled 
out at the 95\% confidence level, a flat model with $\Omega_0 < 0.37$ is ruled out at the same level,   conclusions which are in clear 
disagreement with previous analyses based 
on the same high redshift sample (H97; Eke et al, 1998). The main differences come
from our higher abundance at low redshift and from the fact that we explicitly take into account the effect of temperature measurement errors 
in the high--redshift sample. Note that Viana and Liddle (1999a) reached values which are consistent with ours. There are a number of 
other issues that differ in these various analyses and it is important 
to check whether these differences can result in significantly different 
values of $\Omega_0$.  Eke et al. (1998) have analyzed various sources of systematics 
and concluded that they are not of critical importance. 
We reach similar conclusions for the effects they have investigated. 
For instance, we have modified the $M-T$ relation 
by $\pm$ 20\%, which changes $\Omega_0$ by $\pm 0.1$. The uncertainty in 
the validity of the Press and Schechter formalism is also significant 
but small. Using the revised version by Governato et al. (1998), we found  values arround 15\% higher. We have checked that our likelihood approach 
is not  biased by applying it to the predicted mean abundance in one model 
($\Omega_0 = 0.85$, $n = -1.85$, 
$\sigma_c = 0.606$) and found that the best fitting model is ($\Omega_0 = 0.85$, $n = -1.75$, 
$\sigma_c = 0.61$). We also have checked that varying the shape of the fitting expression does not change the final likelihood at any appreciable level.
After this work was 
finished, we learned about Donahue \& Voit (1999) and Henry (2000) works. The first authors  found a significant evolution in the abundance of clusters over the redshift range [0.-0.8], in agreement 
with Blanchard \& Bartlett (1998) and the present work. Nevertheless, they conclude it to be consistent with a lowdensity universe, but simultaneously they found a very flat spectrum $n \sim -2.2$, which is 
excluded in our analysis based on the 
local temperature distribution function. Similarly, Henry (2000) found a moderate value of $\Omega_0 \sim 0.5 $ using the HA sample. Indeed, the 
strongest source of 
systematic error we found comes from the reference sample we used at 
zero redshift: using the abundances inferred from the sample limited at 
the bright end ($f_x > 4. 10^{-11}$ erg/s/cm$^2$) leads to $\Omega_0 = 0.65$ consistent with  Eke et al.(1998) and Henry (2000). With the fainter part of our sample 
($2.2 10^{-11}$ erg/s/cm$^2$ $<f_x < 4. 10^{-11}$ erg/s/cm$^2$) we obtained 
$\Omega_0 = 0.98$. The difference obtained just by dividing the sample 
into two statistically equivalent sub-samples is surprisingly large (althouhg one may not worry of a difference at  1.5 $\sigma$ level): 
one would expect the uncertainty to come primarily from the 
high--redshift sample (which comprises only nine clusters) and not from 
the low--redshift sample.\\

Our best fit value of $\Omega_0$ inferred from the H97 sample is significantly 
higher than previous estimates from the same sample but it is consistent with the latest optical result by Borgani et al. (1999). 
 It is therefore important to examine the 
robustness of our analysis. This disagreement with previous results can be due to a higher local abundances of X-ray clusters, and a somewhat different treatment of 
the bias introduced by the errors at high redshift. As we have already argued, it seems 
unlikely that our local sample leads to a significant overestimation of 
the local temperature distribution function. In fact, one could argue that 
we may underestimate the actual $N(T)$, because of the possible 
incompleteness of our sample. However, regarding the agreement between 
our local temperature distribution function and what can be inferred from 
the luminosity function this is rather unlikely. 
The high redshift sample might be more worrisome. As we have seen, a  
systematic difference may exist between two samples, leading to larger
differences in $\Omega_0$ than expected from Poisson noise. It is therefore 
conceivable that the high redshift sample is a statistical fluke. For 
instance, Eke et al. noticed that the temperature distribution within the 
redshift bin [0.3--0.4] in the original Henry's sample is statistically 
surprising as 9 of the clusters lie in the range  [0.3--0.35] (now 8 are left). Another 
possibility is that high--redshift 
clusters are more massive than what one would infer from their apparent 
temperature (averaged over the luminosity which comes essentially from 
the core region). This could be possible for instance, if high--redshift 
clusters are more often dominated by cooling flows (or if the selection 
procedure favors cooling flow clusters), making them appear 
cooler than they actually are and producing an apparent evolution in 
the $M-T$ relation. 
%The discrepancy (if real) between X-ray and lensing masses observed for some %clusters seem to confirm such effect. 
At the same time, in order not to produce an evolution in the $L  -T$ 
relation (not seen in the data),  these high-z clusters would have to 
be fainter. This is not very compelling, since cooling flows are expected 
to increase the X-ray luminosity.

Another potential problem in the above determination of the density 
parameter is the quality of the EMSS sample: if the selection function is 
not well understood, it could be that the sample is missing a significantly
larger fraction of the cluster population than expected.  Indeed, clusters 
selection in EMSS is rather problematic, as the detection algorithm 
was designed to detect point sources. A mean correction for extended 
sources was applied (see Gioia and Luppino, 1994 and OB97 for more details), 
but one may nonetheless worry that this procedure is not well controlled.
However, the deficit of high-z clusters observed in Figure \ref{fig:nt2} 
is of the order of 2 to 3. The possibility that the EMSS selection procedure 
could have missed clusters to such an extent seems unlikely. Furthermore, 
the modelling by OB97 {\em predicted } number counts which are 
subsequently seen to be in good agreement with available ROSAT  
counts argues against significant incompleteness of the EMSS sample. 
An other potential problem 
could be a systematical bias in the EMSS fluxes. No evidence has been found 
by Nichol et al. (1997). However, Ebeling et al. (1999) claim that a 
40\% offset in flux exists, which could explain half of the observed dimming. 
This possibility does not seem very appealing, because it would imply a 
very significant evolution in the $L_x-T_x$ relation. In order to get rid of 
possible limitations of the EMSS sample, it will clearly be very 
important to see whether consistent results could be obtained from 
ROSAT selected samples of X-ray clusters. 

\section{Conclusion}

The local temperature distribution function of X-ray clusters is an important tool 
for cosmology and can provide direct information on the dark matter 
distribution. It is therefore essential to have a good estimation of the 
local temperature distribution function. Furthermore, such a sample is
of crucial importance as a reference sample in order to properly evaluate 
the evolutionary properties of the 
cluster population and for other cosmological applications. 
We have provided a new estimate of the temperature 
distribution function for local clusters, based on a large sample of X-ray 
clusters with measured temperatures. This sample is essentially flux limited and we have argued that it is likely to be reasonably representative with a completeness estimated to 85\%. We found an appreciably higher 
abundance of clusters than previous estimates, but in good agreement with 
the abundance inferred from the luminosity function and optical data. We have used the
sample to study the statistical properties of the matter density 
fluctuations and obtained results consistent with previous works, although 
we obtained a normalization slightly higher, $\sigma_8\sim 0.6$ for $\Omega_0 = 1.$ than previously found from X-ray clusters due to our 
higher abundance of 4 keV clusters. Probably the most important application 
of this new temperature function concerns the determination of the density 
parameter via the test of the evolution of the cluster abundance with redshift. 
In order to apply this test, we have used the Henry's sample
which provides for the first time a direct estimation of the temperature 
distribution function at a non-zero redshift. We have found a clear 
indication that the abundance of clusters was smaller at the epoch 
corresponding to $z \sim 0.33$ consistent with Donahue and Voit (1999).
Using a likelihood approach, we inferred a high value of $\Omega_0 \sim 0.9 
\pm 0.2$. Low--density open universes ($\Omega_0 \leq 0.49$) are excluded at 
the 2 $\sigma$ level. The exclusion region is nearly as severe in 
the flat case: $\Omega_0 \leq 0.37$ models are excluded at the 2 $\sigma$ level. 
This result is entirely consistent with other independent analyses of the 
EMSS sample (Sadat et al., 1998; Reichart et al., 1999). We therefore 
confirm that the abundance of X-ray clusters as inferred from the EMSS 
favors a high--density universe. We have pointed out that this result is 
also consistent with what is known from existing ROSAT samples, although 
the situation is not as clear as in the case of the EMSS sample. It is 
important to keep in mind that the present method is one of the very few 
cosmological probes of $\Omega_0$ that is 
not based on local estimates, but is rather global in nature. 
However, given the importance of the conclusion, we believe that it should 
still be considered with caution. Our study of the local temperature 
distribution function demonstrated that systematic uncertainties could 
be more important than expected. It is therefore essential to perform this 
test using an entirely different and independent sample. Temperature measurements with XMM of ROSAT 
selected clusters will allow to obtain such a sample. The application of the cosmological test will then probably lead to a more definitive conclusion concerning
the mean density of the universe.


\begin{thebibliography}{}
\bibitem {}  Arnaud, M. 1994, Cosmological Aspects of X-ray Clusters of Galaxies, W.C.Seitter ed., NATO ASI Series, Vol. 441, 197
\bibitem {}  Arnaud, M. \& Evrard, A.E.  1999, MNRAS, 305, 661
\bibitem{} Bahcall, N.A. \& Fan, X., 1998, ApJ, 504, 1
\bibitem {} Balland, C. \& Blanchard, A. 1997, ApJ, 497, 541 
\bibitem {} Barbosa D., Bartlett J.G., Blanchard A. \& Oukbir, J. 1996, A\&A,
314, 13  
\bibitem {} Bartlett, J.G. 1997, Proceedings of the 
1st Moroccan School of Astrophysics, 
ed. D.~Valls-Gabaud et al., A.S.P. Conf. Ser.,
vol. 126, p. 365
\bibitem {} Bartlett, J.G. \& Silk, J. 1993,  ApJL, 407, L45
\bibitem {} Blanchard, A. \& Bartlett, J. 1998, A\&A, 314, 13  
\bibitem {} Blanchard, A., Bartlett, J. \& Sadat, R.  1999, CRAS, 327, 318. 
\bibitem {} Blanchard, A. \& Silk, J. : 1991,
proceedings of the  Moriond Conference, 1991, Editions Fronti\`eres, p93.
\bibitem {} Bryan,  G.L. \& Norman, M.L. 1998a, ApJ, 495, 80 
\bibitem {} Bryan,  G.L. \& Norman, M.L.  1998b, astro-ph/9802335
\bibitem {} Borgani, S., Rosati, P., Tozzi, P. \& Norman, C. 1999, ApJ, 517, 40.
\bibitem {} Borgani, S., Girardi, M.,  Carlberg, R.G.,  Yee, H.K.C. \&  Ellingson,  E. 1999, ApJ, 527, 561
\bibitem {}  Carlberg, R.G., Morris, S.L., Yee, H.K.C. \& Ellingson, E. 1997,
ApJ, 479, L19  
\bibitem {} Colafrancesco, S., Mazzotta, P. \& Vittorio, N. 1997, ApJ, 488, 566 
\bibitem {} David et al 1993
\bibitem {}  De Grandi, S. et al.  1999a, ApJL, 513, L17  
\bibitem{}  De Grandi, S.  et al. 1999b, ApJ, 514, 148 
\bibitem {}  Donahue, M. 1996, ApJ, 468, 79
\bibitem {}  Donahue, M., Voit, G. M., Gioia, I., Lupino, G., Hughes, J. P. \&
 Stocke, J. T. 1998, ApJ, 502, 550
\bibitem {}  Donahue, M., Voit, G. M., Scharf, C. A., Gioia, I.,  
Mullis, C. P., Hughes, J. P. \&  Stocke, J. T. 1999, ApJ, 527, 525
\bibitem {}  Donahue, M. \& Voit, G. M. 1999, ApJL, 523, L137 
\bibitem {}  Ebeling, H., Voges, W., B\"ohringer, H., Edge, A.C., Huchra, J.P. \& Briel, U.G. 1996, MNRAS, 281, 799
\bibitem {}  Ebeling, H.,   Edge, A.C., Fabian,  A.C.,  Allen, S.W.,  
Crawford, C.S. \& B\"ohringer, H. 1997, ApJL, 479, L101 
\bibitem {}  Ebeling, H.,   Edge, A.C., B\"ohringer, H., Allen, S.W.,  
Crawford, C.S. Fabian,  A.C.,  Voges, W. \& Huchra, J.P. 1998, MNRAS, 301, 881
\bibitem {}  Ebeling, H., Jones, L. R.,   Perlman, E., Scharf, C.,  Horner, D.,
 Wegner, G.,  Malkan,  M., Fairley, B.C. \&  Mullis R. 2000, Ap.J., 534, 133
\bibitem {}  Edge, A.C., Stewart, G.C., Fabian, A.C., \& Arnaud, K.A. 1990,
 MNRAS, 245, 559
\bibitem {} Eke, V.R., Cole, S.,  Frenk, C.S., Henry, P.J. 1996,
MNRAS, 298, 1145
\bibitem {} Eke, V.R., Cole, S.,  Frenk, C.S., Henry, P.J. 1998,
MNRAS, 298, 1145
\bibitem {} Evrard, A.E.  1989,  ApJ, 341, L71
\bibitem {} Evrard, A.E., Metzler, C.A., Navarro, J.F. 1996,  ApJ, 469, 494
\bibitem {} Fabian, A.C.  Crawford, C. S.,  Edge, A. C., Mushotzky, R. F. 
1994, MNRAS, 267, 779
\bibitem {} Frenk, C.S., White, S.D.M., Efstathiou, G., \& Davis, M. 1990,
ApJ, 351, 10
\bibitem {} Frenk, C.S. et al. 1999, ApJ, 525, 554  
\bibitem {} Fukazawa, Y., Makishima, K., Tamura, T., Ezawa, H., Xu, H,Ikebe, Y., Kikushi,K., Ohashi, T. 1998, PASJ, 50, 187.
\bibitem {}  Gioia, I.M., Luppino, G.A. 1994, ApJS, 94, 583  
\bibitem{}  Girardi M., Borgani S., Giuricin G., Mardirossian F. Mezzetti, M. 1998, ApJ, 506, 45
\bibitem{}  Governato F., Babul, A.,  Quinn, T., Tozzi, P.,  Baugh, C. M., 
 Katz, N.\&  Lake, G. 1999, MNRAS, 307, 949
\bibitem {}  Hattori, M. \&  Matsuzawa, H. 1995, A\&A, 300, 637
\bibitem {}  Henry, J.P. \& Arnaud, K.A. 1991, ApJ, 372, 410
\bibitem {}  Henry, J.P.  1997, ApJ, 489, L1
\bibitem {}  Henry, J.P.  2000, ApJ, 535, 350
\bibitem {}  Hughes, J.P., Butchler, J.A., Stewart, G.C. \& Tanaka, Y. 1993, ApJ, 404, 611
\bibitem {} Johnstone, R.M., Fabian, A.C. \& Taylor, G.B. 1998, MNRAS, 298, 854
\bibitem {}  Kaiser, N, 1986,  MNRAS, 222, 323
\bibitem {}  Kruse, G. \&   Schneider P. 1999, MNRAS, 302, 821. 
\bibitem {} Lacey, C. \& Cole, S. 1994, MNRAS 271, 676
\bibitem {} Le Dour, M.,  Bartlett, J.G., Douspis, M. \& Blanchard, A.  2000, astro-ph/0004283, submitted to A\&A 
\bibitem {} Lineweaver, C., Barbosa, D., Blanchard, A. \& Bartlett, J. 1997, 
A\&A, 322, 365
\bibitem {} Lineweaver, C. \& Barbosa, D. 1998, ApJ,  496, 624
\bibitem {} Markevitch, M. 1998, ApJ, 504, 27
\bibitem {} Metzler, C.A. \&   Evrard, A.E. 1997, astro-ph/9710324
\bibitem {}  Mushotzky R. F. \& Scharf C. A. 1997,  ApJ, 482, L13
\bibitem {} Navarro, J. F., Frenk, C.S. \&  White, S.D.M. 1995,
MNRAS, 275, 720
\bibitem {} Nevalainen, J., Markevitch, M., Forman, W.R. 2000, Ap.J., 532, 694
\bibitem {} Nichol, R.C., Holden, B.P., Romer, A.K., Ulmer, M.P., 
Burke, D.J. \& Collins, C.A. 1997, ApJ., 481, 644
\bibitem {} Perrenod, S.C. 1980, ApJ, 236, 373
\bibitem {}  Oukbir, J. \&  Blanchard A. 1992, A\&A, 262, L21   
\bibitem {}  Oukbir, J. \&  Blanchard A. 1997, A\&A, 317, 10   
\bibitem {}  Oukbir, J., Bartlett, J.G. \&   Blanchard, A.  1997, A\&A, 320, 365
\bibitem {}  Peebles, P. J. E., Daly, R. A. \& Juszkiewicz, R. 1989, ApJ, 347, 563
\bibitem {}  Press, W.H. \& Schechter, P. 1974, ApJ, 187, 425 
\bibitem {}  Reichart, D.E.  et al, 1999, ApJ, 518, 521
\bibitem {} Robinson, J., Gawiser, E. \& Silk, J. 2000, Ap.J., 532, 1
\bibitem {}  Roussel, H., Sadat, R., Blanchard, A.  2000, A\&A, submitted
\bibitem {}  Sadat, R., Blanchard, A. \& Oukbir, J. 1998, A\&A,  329, 21
\bibitem {} Schmidt, M. 1968, ApJ, 151, 393
\bibitem {} Viana, P.T.R. \& Liddle,  A.R. 1996, MNRAS, 281, 323
\bibitem {} Viana, P.T.R. \& Liddle,  A.R. 1999a, MNRAS, 303, 535
\bibitem {} Viana, P.T.R. \& Liddle,  A.R. 1999b, electronic proceedings of the conference ``Cosmological Constraints from X-ray Clusters'', Strasbourg,
France, Dec. 9-11, 1998, http://astro.u-strasbg.fr/amas/proc/proceed.html
\bibitem{} Voit, G.M. \& Donahue, M. 1998, ApJ, 500, L111 
\bibitem{} White, S.D.M., Navarro, J.F., Evrard, A.E. \& Frenk, C.S. 1993, 
Nature, 366, 429
\bibitem{} Yamashita, K., in Frontiers of X-ray astronomy, Ed. Y. Tanaka, K., Koyama (Universal Academy Press, Tokyo), p.475
\end{thebibliography}
\end{document}